\documentclass[physics,article,accept,moreauthors,pdftex]{Definitions/mdpi}

\newcommand{\dd}{\textrm{d}}

\firstpage{1} 
\pubvolume{1}
\issuenum{1}
\articlenumber{0}
\pubyear{2021}
\copyrightyear{2020}
\datereceived{} 
\dateaccepted{} 
\datepublished{} 
\hreflink{https://doi.org/} % If needed use \linebreak
\Title{DETERMINING PITCH-ANGLE DIFFUSION COEFFICIENTS  FOR ELECTRONS IN WHISTLER TURBULENCE}

\TitleCitation{Determining pitch-angle diffusion coefficients  for electrons in whistler turbulence}

\Author{Felix Spanier $^{1}$, Cedric Schreiner $^{2,3}$ and Reinhard Schlickeiser $^{4,5}$}

\AuthorNames{Felix Spanier, Cedric Schreiner and Reinhard Schlickeiser}

\AuthorCitation{Spanier, F.; Schreiner, C.; Schlickeiser, R.}
\address{%
$^{1}$ \quad Institut f\"ur Theoretische Astrophysik, Universit\"at Heidelberg, Albert-Ueberle-Strasse 2, 69120 Heidelberg, Germany; felix.spanier@uni-heidelberg.de\\
$^{2}$ \quad Centre for Space Research, Northwest-University,Potchefstroom, South Africa;
mail@cschreiner.de\\
$^{3}$ \quad Max-Planck-Institute for Solar System Research, Justus-von-Liebig-Weg 3, 37077 G\"ottingen, Germany\\
$^{4}$ \quad Institut f\"ur Theoretische Physik IV, Ruhr-Universit\"at Bochum, Universit\"atsstrasse 150, 44801 Bochum, Germany; rsch@tp4.rub.de\\
$^{5}$ \quad Institut f\"ur Theoretische Physik und Astrophysik, Christian-Albrechts-Universit\"at zu Kiel, Leibnizstr. 15, D-24118 Kiel, Germany}

\corres{Correspondence: felix.spanier@uni-heidelberg.de}

\abstract{Transport of energetic electrons in the heliosphere is governed by resonant interaction with plasma waves, for for electrons with sub-GeV kinetic energies specifically with dispersive modes in the whistler regime. We have performed Particle in Cell simulations of kinetic turbulence with test-particle electrons. The pitch-angle diffusions coefficients of these test-particles have been analyzed and compared to an analytical model for left- and right-handed polarized wavemodes.}


\begin{document}
%%%%%%%%%%%%%%%%%%%%%%%%%%%%%%%%%%%%%%%%%%

%
%
%\markboth{Spanier, Schreiner \& Schlickeiser}
%{Pitch-angle scattering in Whistler turbulence}
%
%%%%%%%%%%%%%%%%%%%%%% Publisher's Area please ignore %%%%%%%%%%%%%%%
%%
%\catchline{}{}{}{}{}
%%
%%%%%%%%%%%%%%%%%%%%%%%%%%%%%%%%%%%%%%%%%%%%%%%%%%%%%%%%%%%%%%%%%%%%%
%
%\title{DETERMINING PITCH-ANGLE DIFFUSION COEFFICIENTS  FOR ELECTRONS IN WHISTLER TURBULENCE  }
%
%\author{F. SPANIER}
%
%\address{Institut f\"ur Theoretische Astrophysik, Universit\"at Heidelberg, Albert-Ueberle-Strasse 2\\
%69120 Heidelberg, Germany\\
%felix.spanier@uni-heidelberg.de}
%
%\author{C. SCHREINER}
%
%\address{Centre for Spherical Rubberducks, Northwest-University\\
%Potchefstroom, South Africa\\
%mail@cschreiner.de}
%
%
%\author{R. SCHLICKEISER}
%
%\address{Institut f\"ur Theoretische Physik IV, Ruhr-Universit\"at Bochum, Universit\"atsstrasse 150\\
%44801 Bochum, Germany\\
%rsch@tp4.rub.de}
%
%\maketitle
%
%
%
%\keywords{Keyword1; keyword2; keyword3.}
%
%\ccode{PACS numbers:}

%\tableofcontents

\section{Introduction}

The solar wind and most phases of the interstellar medium are strongly 
turbulent media with high \emph{magnetic Reynolds numbers} of 
$10^{14}$\cite{borovsky_2003}. Turbulence manifests 
itself in a spectrum of plasma waves at various length scales and frequencies.
The energy distribution as a function of the frequency follows a 
characteristic power law. The current understanding of the turbulent 
processes is such that energy is injected at large scales (i.e.~small wave 
numbers and frequencies) and then cascades to smaller spatial scales.

The energy spectrum can be divided into several regimes, which each may span 
several orders of magnitude in wave number or frequency.
At largest scales the \emph{injection range} is found, which then transitions 
into the \emph{inertial range}. The inertial range can be described by 
magnetohydrodynamic (MHD) theory and turbulence is dominated by the 
interaction of Alfv\'en waves. At smaller scales kinetic effects of the 
particles come into play. 

This high wave number regime is often referred to as kinetic, dispersive, or dissipation range of the  spectrum, since the waves become dispersive 
and dissipation starts to set in. While the spectrum extends to even smaller 
scales, damping eventually becomes dominant and leads to an exponential 
cutoff of the energy spectrum.

Power law distributions of the fluctuating magnetic energy are expected in 
the injection, inertial, and dissipation range of the spectrum. However, 
the spectral indices may differ among the individual regimes. The work of 
Goldreich and Sridhar \cite{sridhar_1994,goldreich_1995}
presents a detailed model of the turbulent energy cascade in the inertial 
regime. Their model predicts a spectral index of $-5/3$, which actually seems 
to be realized in the solar wind \cite{2010PhRvL.105m1101S}.
Subsequent models by Galtier et al. \cite{2000JPlPh..63..447G,2002ApJ...564L..49G} 
give rise to a spectral index $k_\perp^{-2}$.

Kinetic turbulence in the dissipation range is an active field of research 
\cite{howes_2015_b}. Especially the composition of the wave spectrum is 
subject to discussion, because a transition from non-dispersive Alfv\'en 
waves to dispersive wave modes is expected. Possible candidates for the waves 
in the dissipation range are so-called \emph{kinetic Alfv\'en} and \emph{whistler waves} \cite{gary_2009}.

The impact of kinetic turbulence on the transport of energetic particles is another major topic.
 %Here, especially the electrons are of interest.
The transport of energetic protons is well-described by models of Alfv\'enic 
turbulence, since the protons mainly interact with these waves at low 
frequencies. The theoretical framework of quasi-linear theory (QLT) can be 
used to describe particle transport by a series of resonant interactions with 
the magnetic fields of Alfv\'en waves, which leads to scattering of the 
particles \cite{jokipii_1966,lee_1974,schlickeiser_1989}. This theory 
describes changes of the particles' pitch angles (i.e.~the angle 
of the velocity vector relative to a background magnetic field), momenta, or 
positions as diffusion processes and allows to predict diffusion coefficients 
and other quantities, such as the mean free path, which can be compared to 
observations.

Dispersive waves are more difficult to handle in (analytical) theory.
Nonetheless, QLT can also yield predictions for particle transport in a 
medium containing dispersive waves\cite{steinacker_1992,vainio_2000}. The 
introduction of dispersive waves can even solve some of the problems which 
are encountered if a purely Alfv\'enic spectrum of waves is assumed 
\cite{achatz_1993}. Still, the model remains an approximation and computer 
simulations are used to clarify some of the details which are not included in 
the analytical theory. Different kinds of simulations are used to study 
different physical regimes and processes, from the acceleration of particles 
\cite{ng_1994,vainio_2001} to the transport of energetic particles, 
considering both non-dispersive \cite{lange_2013} and dispersive waves 
\cite{gary_2003,camporeale_2015}.

The key problem which has been chosen for the subject of this work is the 
process of electron transport in dispersive turbulence. The transport of
electrons at sub-GeV energies has been of high interest for quite some time\cite{1998GeoRL..25.4099M}. 
As was 
mentioned in the previous sections, particle acceleration in  Alfv\'enic 
turbulence in the inertial range is well-understood.
However, turbulence on kinetic scales still poses problems for both 
observations and modeling.

\section{Theory}

\subsection{Turbulence theory}
\label{sec:turbtheory}

From the observations it is not entirely clear which types of plasma waves 
constitute the spectrum of turbulent waves in the dispersive and dissipative 
regime. \emph{Kinetic Alfv\'en} and \emph{whistler waves} are both possible 
candidates\cite{gary_2009}. While kinetic Alfv\'en waves (KAWs) simply 
represent the continuation of the Alfv\'en mode (Eq. \eqref{eq:omegal}) in 
the dispersive regime with very large perpendicular wave numbers (perpendicular 
wavelength in the range of proton gyroradius), whistler waves 
(Eq. \eqref{eq:omegar}) are right-handed polarized modes at large wave numbers. 
It is perfectly reasonable to assume that non-dispersive Alfv\'en waves 
simply transition to KAWs. However, observations reveal that whistler waves 
are also present in various regions of the heliosphere, such as in the 
interplanetary medium \cite{droege_2000}, close to interplanetary shocks 
\cite{coroniti_1982,aguilar-rodriguez_2011} or planetary bow shocks 
\cite{fairfield_1974}, and also in the Earth's ionosphere and foreshock region 
\cite{kennel_1966,palmroth_2015}.

Whereas left-handed polarized Alfv\'en waves are damped by protons and cannot 
cascade to frequencies above the proton cyclotron frequency, a spectrum of 
whistler waves may extend to frequencies beyond the ion-cyclotron frequencies. 
Whistler waves primarily interact with electrons and are also damped by 
electrons at higher frequencies (close to the electron cyclotron frequency). 
This is an interesting aspect of kinetic turbulence, since a population of 
whistler waves can heat the electrons in the solar wind or even accelerate 
particles in the high energy tail of the thermal spectrum.

\emph{Kinetic turbulence} includes the smallest length scales, where the 
interaction of waves and particles becomes important. Although the 
wave-particle interactions are not explicitly included in theory, their effect has 
to be considered by allowing dispersive waves and damping. This regime is 
generally more complicated and less well understood than MHD turbulence.

The general picture associated with turbulence is the following: Energy is 
injected into the system at a large outer scale (small wave numbers $k$). The 
energy is transported via the interaction of waves to smaller spatial scales 
(larger wave numbers) and the (magnetic) energy spectrum $E_B(k)$ follows a 
power law distribution. This is the inertial range. The spectrum steepens as 
the kinetic regime or \emph{dissipation range} is reached, but energy is 
still transported to smaller scales. First, only ion effects will start 
influencing the plasma dynamics, but at even larger wave numbers the 
electrons can also interact with the plasma waves. This is where the energy 
spectrum is cut off. One aspect that has been discussed in greater detail since 
the seminal work of Goldreich and Sridhar\cite{goldreich_1995} is the possible 
anisotropy with respect to the background magnetic field: The turbulent
spectrum may behave differently depending on wave numbers parallel ($k_\parallel$) 
or perpendicular ($k_\perp$) to the background magnetic field.

The special case of whistler turbulence has been discussed in greater detail. 
The properties of this whistler turbulence have been analyzed using kinetic 
simulations in two \cite{gary_2008,che_2014} and three dimensions 
\cite{gary_2012,chang_2013,gary_2014,chang_2015}.
These studies suggest a steeper energy spectrum than for Alfv\'enic 
turbulence, with a spectral index $\sigma$ in the range between $-3.7$ and 
$-5.5$ and a possible break in the energy spectrum \cite{chang_2015}.
Results by Chang\cite{chang_2013} also suggest that the wave vector 
anisotropy depends on the choice of the plasma beta.
A relatively isotropic spectrum is obtained for a plasma beta $\beta \sim 1$, whereas 
$\beta < 1$ yields an anisotropic cascade which favors the transport of energy to 
larger $k_\perp$. The plasma beta is the ratio of thermal to magnetic energy.
 The anisotropy additionally depends on the energy deployed 
to the electromagnetic fields of the turbulent whistler waves \cite{gary_2012}.

\subsection{Subluminal parallel waves in cold plasmas}

\label{sec:disprel}

In warm thermal plasmas with low plasma betas the real part of the dispersion relation
agrees very well with the cold plasma dispersion relation, so that we use the latter here. We also ignore here the resonance broadening effects caused by a finite imaginary part
of the dispersion relation in warm plasmas implying a finite weak-damping rate; these effects were considered in Schlickeiser and Achatz\cite{1993JPlPh..49...63S}.

Using the convention of positive frequencies $\omega > 0$, where $\omega$ is the real part of the generally complex frequency, but parallel wave numbers $k_\parallel \in [-\infty,\infty]$ (here the case $k_\perp = 0$, also known as slab case, is treated) the dispersion relation of right-(R) and left-handed (L) polarized undamped low-frequency ($\omega \leq \Omega_{e,0}$) parallel Alfv\'en wave in a cold electron-proton background plasma  reads \cite{swanson89}

\begin{align}
	n_L^2 &= 1 - \frac{\omega_{pi}^2}{\omega(\omega-\Omega_i)} - \frac{\omega_{pe}^2}{\omega(\omega + \Omega_e)},\\
	n_R^2 &= 1 - \frac{\omega_{pi}^2}{\omega(\omega+\Omega_i)} - \frac{\omega_{pe}^2}{\omega(\omega - \Omega_e)},\\
	\frac{k_\parallel^2 c^2}{\omega_{L,R}^2} &=  1 - \frac{\omega_{pi}^2}{\omega(\omega\mp\Omega_i)} - \frac{\omega_{pe}^2}{\omega(\omega \pm \Omega_e)},\\
	\frac{k_\parallel^2 c^2}{\omega_{L,R}^2} - 1&=-\frac{c^2\Omega_i^2}{v_A^2} \frac{M+1}{(\omega\mp\Omega_i)(\omega\mp M \Omega_i)} \label{eq:rsch_disprel}
\end{align}
with the mass ratio $M=m_p/m_e=1836$ and the Alfv\'en speed $v_A=\beta_A c = 2.18\times 10^{11} B [\text{G}] n_i^{-1/2} [\text{cm}^{-3}] \text{cm/s}$. For subluminal wave phase speeds
\begin{align}
	|V_\text{phase}| = |\frac{\omega_{L,R}}{k_\parallel}| \ll c,
\end{align}
which has to be checked a posteriori, the dispersion relation \eqref{eq:rsch_disprel} simplifies to
\begin{equation}
		\frac{k_\parallel^2 c^2}{\omega_{L,R}^2} \simeq - \frac{(M+1)\Omega_i^2 \omega_{L,R}^2}{v_A^2(\omega\mp\Omega_i)(\omega\mp M \Omega_i)}.
\end{equation}
It is convenient to introduce dimensionless frequencies and parallel wave numbers by defining
\begin{equation}
	y_{L,R} = \frac{\omega_{L,R}}{\Omega_i}>0, \quad k = \frac{k_\parallel}{k_c}
	\label{eq:dimensionless}
\end{equation}
with
\begin{equation}
	k_c = \frac{\Omega_i}{v_A}
\end{equation}
so that the subluminal dispersion relation becomes
\begin{equation}
	k^2 = -\frac{(M+1) y_{L,R}^2}{(y_{L,R}\mp 1)(y_{L,R}\pm M)}
	\label{eq:k2disp}
\end{equation}
with the two solutions
\begin{equation}
	k_{1,2} = -\frac{\sqrt{(M+1)} y_{L,R}}{\sqrt{(y_{L,R}\mp 1)(y_{L,R}\pm M)}}.
\end{equation}
As $y_{L,R}$ is always positive, the positive solution $k_1>0$ describes forward moving waves with positive phase speed, whereas the negative solution $k_2=-k_1 <0$ describes backward moving waves.

\subsubsection{Left-handed modes}
Equation \eqref{eq:k2disp} indicates that no left-handed polarized solution with $y_L>1$ exists, so that a further simplification of Eq. \eqref{eq:k2disp} for left-handed polarized waves is possible, assuming that $M\gg 1$  
\begin{equation}
	k^2 \simeq \frac{y_L^2}{1-y_L}
	\end{equation}
with the solutions
\begin{equation}
	y_L(k) =\frac{k^2}{2}\pm\sqrt\frac{k}{2}\sqrt{k^2-4}\simeq\begin{cases}|k| & \text{for}\ k\ll 1\\ 1-\frac{1}{k^2} & \text{for}\ k\gg 1 \end{cases}
	\label{eq:rsch48}
\end{equation}
corresponding to
\begin{equation}
	\omega_L \simeq \begin{cases}  V_A|k_\parallel| & \text{for}\ k_\parallel \ll k_c\\ \Omega_i \left(1-\frac{k_c^2}{k_\parallel^2}\right) & \text{for}\  k_\parallel \gg k_c  \end{cases}
	\label{eq:omegal}.
\end{equation}

\subsubsection{Right-handed modes} The right-handed solutions of Eq. \eqref{eq:k2disp}
\begin{equation}
	k^2 = -\frac{(M+1)y_R^2}{(y_R+1)(y_R+M)}
\end{equation}
 can be approximated under the assumption that $M^{-1}\ll 1$
\begin{equation}
	y_R = \frac{(M+1) k^2}{2(1+k^2+M)}\left( 1-\frac{2}{M+1} + \sqrt{1+
	\frac{4 M}{k^2(M+1)}} \right).
	\label{eq:rsch50}
\end{equation}
Depending on $k$ different regimes can be identified
\begin{equation}
	\omega_R = \begin{cases}
	V_A |k_\parallel| & |k_\parallel|< k_c\\
	\Omega_i + \frac{V_A^2 k_\parallel^2}{\Omega_i} & k_c \leq |k_\parallel | \leq M^{1/2} k_c\\
	\Omega_e\left( 1- \frac{M k_c^2}{k_\parallel^2} \right) & |k_\parallel|>M^{1/2} k_c
	\end{cases}.
	\label{eq:omegar}
\end{equation}
The first range describes the linear dispersion regime, the second the whistler regime and the last range is the electron-cyclotron range. While these approximate solutions are providing good estimates to the real solution they have one major problem: the solutions do not provide continuous coverage. An alternative approximation is
\begin{equation}
	y_R(k) \simeq |k|(1+|k|). 
	\label{eq:rdispsimple}
\end{equation}

In the following we will ignore the particle scattering by parallel waves at electron or ion cyclotron frequencies, as these are highly damped in a realistic warm thermal plasma, so that the resonant interaction does not apply (see \cite{1993JPlPh..49...63S} for a discussion of wave-particle interactions with damped waves). For left-handed and right-handed polarized waves this restricts the normalized wave numbers to values of $k\leq 1$ and $k \leq M$ respectively.

\subsection{Particle transport}

Any charged particle of given velocity $v$, Lorentzfactor $\gamma = (1-(v^2/c^2))^{-1/2}$, pitch angle cosine $\mu=v_\parallel/v$, mass $m$, charge $q_i=e|Z_i| Q$ with $Q=\text{sgn}(Z_i)$ and relativistic gyrofrequency $\Omega^\prime  = Q \Omega/\gamma$ with positive $\Omega= |q| B_0/(mc)$ interacts with parallel right- and left-handed polarized plasma waves whose wave number $k$,
 %cosine of propagation angle $\eta=k_\parallel/k$ 
and real frequency $\omega_{R,L}$ obey the resonance condition
\begin{equation}
v \mu k_\parallel - \omega_{R,L}(k_\parallel)\mp\frac{Q\Omega}{\gamma}=0.
\label{eq:resonance_condition}
\end{equation}
Introducing
\begin{equation}
	x= \frac{p}{m_e c}, \quad \epsilon = \frac{v_A}{v} = \beta_A \frac{(1+x^2)^{1/2}}{x}
\end{equation}
and the dimensionless frequency and wave number \eqref{eq:dimensionless} the resonance condition \eqref{eq:resonance_condition} reads
\begin{equation}
	\Omega\left( \frac{\mu k}{\epsilon} - y_{R,L}(k) \mp S_i \right) = 0
\end{equation}
with 
\begin{equation}
	S_i = \frac{Q |Z_i| m_p}{m \gamma} = \frac{1}{\gamma}\begin{cases}1 & \text{for protons} \\ -M & \text{for electrons} \end{cases}
\end{equation}

The quasilinear Fokker-Planck coefficients for the pitch angle cosine $\mu$ are given by
\begin{eqnarray}
	D_{\mu\mu}(\mu) = \frac{\pi^2 \Omega^2(1-\mu^2)}{B_0^2}\int_{-\infty}^\infty \dd k_\parallel \\\nonumber
	\times \left( g_R(k_\parallel) \delta (v k_\parallel-\omega_R-\Omega^\prime) \left( 1-\frac{\mu\omega_R}{k_\parallel v} \right)^2 \right. + \\
	\nonumber\left. + g_L(k_\parallel) \delta (v k_\parallel-\omega_L+\Omega^\prime) \left( 1-\frac{\mu\omega_L}{k_\parallel v} \right)^2\right)
	\label{eq:fp}
\end{eqnarray}
with the magnetic fluctuation spectra of right- and left-handed polarized waves $g_{R,L}(k_\parallel)$, where the total magnetic field fluctuations are determined by \cite{schlickeiser_2003} 
\begin{equation}
	(\delta B)^2 = 2\pi \int_{-\infty}^\infty \dd k_\parallel(g_L(k_\parallel) + g_R(k_\parallel)).
	\label{eq:rsspektrum}
\end{equation}
In Eq. \eqref{eq:fp} the frequencies $\omega_{R,L}(k_\parallel)$ are determined by the solutions of the dispersion relations discussed in Sec. \ref{sec:disprel}.

In terms of the normalized wave number $k_\parallel = k_c k$  and frequency $\omega_{R,L} = \Omega_i y_{R,L}$ the Fokker-Planck coefficients \eqref{eq:fp} read 
\begin{eqnarray}
	D_{\mu\mu}(\mu) = \frac{\pi^2 \Omega^2 k_c (1-\mu^2)}{\Omega_i^2 B_0^2}\int_{-\infty}^\infty \dd k_\parallel \\\nonumber
	\times \left( g_R(k_\parallel) \delta (\frac{k \mu}{\epsilon}-y_R(k)-S_i) \left( 1-\frac{\mu\omega_R}{k_\parallel v} \right)^2 \right. + \\
	\nonumber\left. + g_L(k_\parallel) \delta (\frac{k \mu}{\epsilon} -y_L+S_i) \left( 1-\frac{\mu\omega_L}{k_\parallel v} \right)^2\right).
	\label{eq:fpnorm}
\end{eqnarray}
The calculation of the Fokker-Planck coefficients requires the knowledge of the magnetic fluctuation spectra. The correct theoretical description is complicated as described in Sec. \ref{sec:turbtheory}, but results from numerical calculations can be inferred.

Deriving the Fokker-Planck coefficients in general is a very involved task, but some of the limiting cases are much simpler to derive. It is helpful to account for the relative abundance of forward and backward moving waves and their polarization state. We  introduce the cross helicities $H_{L,R} \in [-1,1]$ for left- and righthanded polarized waves to write the spectra 
\eqref{eq:rsspektrum} of backward and forward propagating waves as
\begin{align}
	g_{L,R}^b &= \frac{1-H_{L,R}}{2} g_{L,R}(k)\Theta(-k),\\
	g_{L,R}^f &= \frac{1+H_{L,R}}{2} g_{L,R}(k)\Theta(k).
\end{align}
The step functions $\Theta(\pm k)$ appear because backward and forward moving waves only occur at negative and positive wave numbers respectively. In general, these cross helicities can depend on the wave number, but throughout this work we adopt isospectral turbulence where $H_{L,R}$ are constants independent of $k$. The magnetic helicity $\sigma(k)\in[-1,1]$ characterizes the relative abundances of left- and right-handed polarized waves
\begin{align}
	g_L(k) &= \frac{1+\sigma(k)}{2}g_\text{tot}(k),\\
	g_R(k) &= \frac{1-\sigma(k)}{2}g_\text{tot}(k).
\end{align}
For parallel plasma waves we know already that $\sigma(y>1)=-1$ as no left-handed polarized waves exist at these normalized frequencies.

Using these definitions the Fokker-Planck coefficient \eqref{eq:fpnorm} can be calculated for the different helicities
\begin{eqnarray}
	D_{\mu\mu}(\mu) = \frac{\pi^2 \Omega^2 k_c (1-\mu^2)}{\Omega_i^2 B_0^2}\int_{-\infty}^\infty \dd k_\parallel g_\text{tot}(k) 
	\times\\
	\nonumber\left( 	
	(1-H_R)(1-\sigma(k)) \delta (\frac{k \mu}{\epsilon} +y_R+S_i) \left( 1+\frac{\epsilon \mu y_R(k)}{k} \right)^2 \right. + \\
		\nonumber\left. + 
		(1-H_L)(1+\sigma(k)) \delta (\frac{k \mu}{\epsilon}+y_R(k)-S_i) \left( 1+\frac{\epsilon \mu y_L(k)}{k} \right)^2 \right. + \\
	\nonumber\left. + 
	(1+H_R)(1+\sigma(k)) \delta (-\frac{k \mu}{\epsilon} +y_R+S_i) \left( 1-\frac{\epsilon \mu y_R(k)}{k} \right)^2 \right. + \\
	\nonumber\left. + 
	(1+H_L)(1-\sigma(k)) \delta (-\frac{k \mu}{\epsilon}+y_R(k)-S_i) \left(1-\frac{\epsilon \mu y_L(k)}{k} \right)^2 	\right).
	\label{eq:fp2}
\end{eqnarray}

\subsubsection{Interactions in the whistler regime}

For frequencies above the ion cyclotron frequency only right-handed waves exist obeying the dispersive whistler mode dispersion relation. As discussed above the turbulent spectrum typically has a much softer spectral index than 2 (theoretical values are in the range of 3 to 6 \cite{2012PhPl...19e5906T,2015ASSL..407..123H}) in this case. 

We consider the case $H_R=-1$ and $\sigma(k)=-1$, only backward moving right-hand polarized waves, which reduces the Fokker-Planck coefficient \eqref{eq:fp2} to
\begin{eqnarray}
	D_{\mu\mu}(\mu) = \frac{\pi^2 \Omega^2 k_c (1-\mu^2)}{\Omega_i^2 B_0^2}\int_{k_0}^\infty \dd k_\parallel g_\text{tot}(k) 
	\times\\
	\nonumber\left( \delta (\frac{k \mu}{\epsilon} +y_R+S_i) \left( 1+\frac{\epsilon \mu y_R(k)}{k} \right)^2 \right).
	\label{eq:fpr}
\end{eqnarray}
The calculation for forwarding moving waves is similar 
\begin{eqnarray}
	D_{\mu\mu}(\mu) = \frac{\pi^2 \Omega^2 k_c (1-\mu^2)}{\Omega_i^2 B_0^2}\int_{k_0}^\infty \dd k_\parallel g_\text{tot}(k) 
	\times\\
	\nonumber\left( \delta (-\frac{k \mu}{\epsilon} +y_R+S_i) \left( 1-\frac{\epsilon \mu y_R(k)}{k} \right)^2 \right).
\end{eqnarray}
With Eq. \eqref{eq:rdispsimple} the resonance condition with positive values of $k$ reads
\begin{align}
	0 = S_i + k\frac{\mu}{\epsilon} + \begin{cases} |k| & \text{for~} k\leq 1\\k^2 & \text{for~} 1\leq k\leq M^{1/2}=43.  \end{cases}
	\label{eq:rsch72}
\end{align}
We prove that this equation for protons and electrons has only one solution $k_r>0$. In the Alfv\'enic wave number range ($k\leq 1$) this is trivial $k_r=-S_i/(1+(\mu/\epsilon))$, which can be positive depending on the signs of $S_i$ and $\mu$.

In the whistler wave number range ($0\leq k \leq 43$), the proof is a bit more involved. Here Eq. \eqref{eq:rsch72} has the two solutions
\begin{align}
	k_1 &= \frac{1}{2}\left( \sqrt{\frac{\mu^2}{\epsilon^2}-4 S_i} - \frac{\mu}{\epsilon} \right),\label{eq:rsch73a}\\
	k_2 &= -\frac{1}{2}\left( \sqrt{\frac{\mu^2}{\epsilon^2}-4 S_i} + \frac{\mu}{\epsilon} \right).
	\label{eq:rsch73b} 
\end{align}
To obtain real valued solutions \eqref {eq:rsch73a} and \eqref{eq:rsch73b}, the condition $\mu^2\geq 4 S_i \epsilon^2$ has to be fulfilled. Assuming that this is fulfilled, we note that for nonnegative values of $\mu\geq 0$ the solution $k_2(\mu\geq 0)<0$ is always negative leaving only one solution for $k_r=k_1(\mu\geq 0)>0$. Alternatively, for negative values of $\mu = -|\mu|$ the solutions \eqref {eq:rsch73a} and \eqref{eq:rsch73b} become
\begin{align}
	k_1(\mu<0) &= \frac{1}{2}\left( \sqrt{\frac{\mu^2}{\epsilon^2}-4 S_i} - \frac{|\mu|}{\epsilon} \right),\label{eq:rsch74a}\\
	k_2(\mu<0) &= \frac{1}{2}\left( \frac{|\mu|}{\epsilon} - \sqrt{\frac{\mu^2}{\epsilon^2}-4 S_i} \right).
	\label{eq:rsch74b}
\end{align}
To further evaluate the solution it is necessary to distinguish between positive and negative values of $S_i$ (i.e., protons and electrons). For electrons ($S_i<0$) the solution $k_2(\mu<0)<0$ is again always negative. For protons ($S_i>0$), both solutions \eqref{eq:rsch74a} and \eqref{eq:rsch74b} are positive, but the second one is always smaller than 
\begin{equation}
	k_2(\mu<0, S_i>0) \leq S_i^{1/2} = \frac{1}{2\gamma} < 1
\end{equation}
as for protons $S_i = \gamma^{-1}$. This solution is positive, but is in contradiction to the previous assumption of modes in the whistler regime with $k>1$. This leaves us with only one solution for the resonant wave number $k_r=k_1(\mu<0,S_i>0)$ in the whistler wave number range.

We then obtain
\begin{align}
	D_{\mu\mu}(\mu) = \frac{\pi^2 \Omega^2 k_c (1-\mu^2)}{\Omega_i^2 B_0^2}\Theta[k_r-k_0]\Theta[M^{1/2}-k_r]  \times \label{eq:fpr2} \\
	\times \frac{g_\text{tot}(k_r)}{|\frac{\dd y_R}{\dd k}+\frac{\mu}{\epsilon}|_{k=k_r}}  \left( 1+\frac{\epsilon \mu y_R(k_r)}{k_r} \right)^2. \nonumber
\end{align}

The case of forward moving right-handed polarized waves is similar to the backward moving waves. The main difference is the resonant wave number
\begin{equation}
	k_r = \frac{1}{2}\left( \sqrt{\frac{\mu^2}{\epsilon^2}-4 S_i} + \frac{\mu}{\epsilon} \right).
\end{equation}

\subsubsection{Alfv\'en and whistler contributions}

The total Fokker-Planck coefficient can be written as
\begin{align}
	D_{\mu\mu} (\mu) = D_{\mu\mu}^\text{A} + D_{\mu\mu}^\text{W}
\end{align}
which is  a sum of Alfv\'en and whistler contributions. For the Alfv\'en wave Fokker-Planck coefficients we insert the asymptotic expansions $y_{R,L} (k \leq 1) \simeq k$ of Eqs. \eqref{eq:rsch48} and \eqref{eq:rsch50} to obtain
\begin{align}
		D_{\mu\mu}^\text{A}(\mu) = 
		\frac{\pi^2 \Omega^2 k_c (1-\mu^2)}{\Omega_i^2 B_0^2}
		\int_{k_0}^1 \dd k_\parallel g_\text{tot}(k) 
		\times 	\label{eq:dmmtotal1} \\
		\left(  \left( 1+\epsilon \mu  \right)^2 
		\left[ (1-H_R)(1-\sigma) 	\delta (k (1+\frac{\mu}{\epsilon})+S_i)+\right.\right.\nonumber\\
		\left.\left. 
			+(1-H_L)(1+\sigma)\delta (k (1+\frac{\mu}{\epsilon})-S_i)	\right]+
		\right.\nonumber\\
		\left. +	\left( 
			1-\epsilon \mu  \right)^2 \left[ (1+H_R)(1-\sigma) 	\delta (k (1-\frac{\mu}{\epsilon})+S_i)
		+ \right.\right. \nonumber\\
	\left.\left. +
		(1+H_L)(1+\sigma) 	\delta (k (1-\frac{\mu}{\epsilon})-S_i)
	\right]
	\right).	\nonumber
\end{align}

The whistler contributions have already been calculated before and are given here for completeness in the same form
\begin{eqnarray}
		D_{\mu\mu}^\text{W}(\mu) = \frac{\pi^2 \Omega^2 k_c (1-\mu^2)}{\Omega_i^2 B_0^2}\int_{1}^{M^{1/2}} \dd k_\parallel g_\text{tot}(k) 
	\times\\
	\nonumber
		\left(  \left( 1+\epsilon \mu  \right)^2  (1-H_R)(1-\sigma) 	\delta (k^2 + k\frac{\mu}{\epsilon}+S_i)
		+\right. \\
	\nonumber
\left. +		\left( 1-\epsilon \mu  \right)^2  (1+H_R)(1-\sigma) 	\delta (k^2 - k\frac{\mu}{\epsilon}+S_i)
	\right).
	\label{eq:dmmtotal2}
\end{eqnarray}

\subsubsection{Electrons}

\label{sec:transport:electron}

The Fokker-Planck coefficients derived in the previous section hold for any particle. In general the integration is simple as the delta-function of the resonance condition yields a simple result. A specific turbulent spectrum has to be defined. We refrain from performing the integral here, but will point out which interaction will take place. There is a clear difference between electrons and protons, we will limit our discussion to the electron case.

For Alfv\'en waves we can distinguish the interaction of electrons with forward/backward moving left-/right-handed modes. Defining 
\begin{align}
	\mu_R(x) &= \frac{\beta_A(M-\sqrt{1+x^2})}{x},\\
	\mu_L(x) &= \frac{\beta_A(M+\sqrt{1+x^2})}{x},
\end{align}
we can constrain the waves for which resonant interaction with electrons is possible:
\begin{enumerate}
	\item backward moving, RH-polarized (bR) Alfv\'en waves for all pitch angle cosines with $\mu\geq \mu_R(x)$ and $\mu \geq -\epsilon = -\beta_A\sqrt{1+x^2}/x$
	\item backward moving, LH-polarized (bL) Alfv\'en waves for all pitch angle cosines with $\mu\leq \mu_L(x)$ and $\mu \leq -\epsilon = -\beta_A\sqrt{1+x^2}/x$
	\item forward moving, RH-polarized (fR) Alfv\'en waves for all pitch angle cosines with $\mu\leq \mu_R(x)$ and $\mu \geq \epsilon = \beta_A\sqrt{1+x^2}/x$
	\item forward moving, LH-polarized (fL) Alfv\'en waves for all pitch angle cosines with $\mu\geq \mu_L(x)$ and $\mu \geq \epsilon = \beta_A\sqrt{1+x^2}/x$
\end{enumerate}

For whistler waves additionally
\begin{equation}
	\mu_0(x) = \frac{\beta_A M^{1/2}(\sqrt{1+x^2}-1)}{x}
\end{equation}
is defined. The resonant interaction takes place between
\begin{equation}
	-\mu_0(x) \leq \mu \leq \mu_R(x).
\end{equation}

\begin{figure}
	\begin{center}
		\includegraphics[width=0.95\columnwidth]{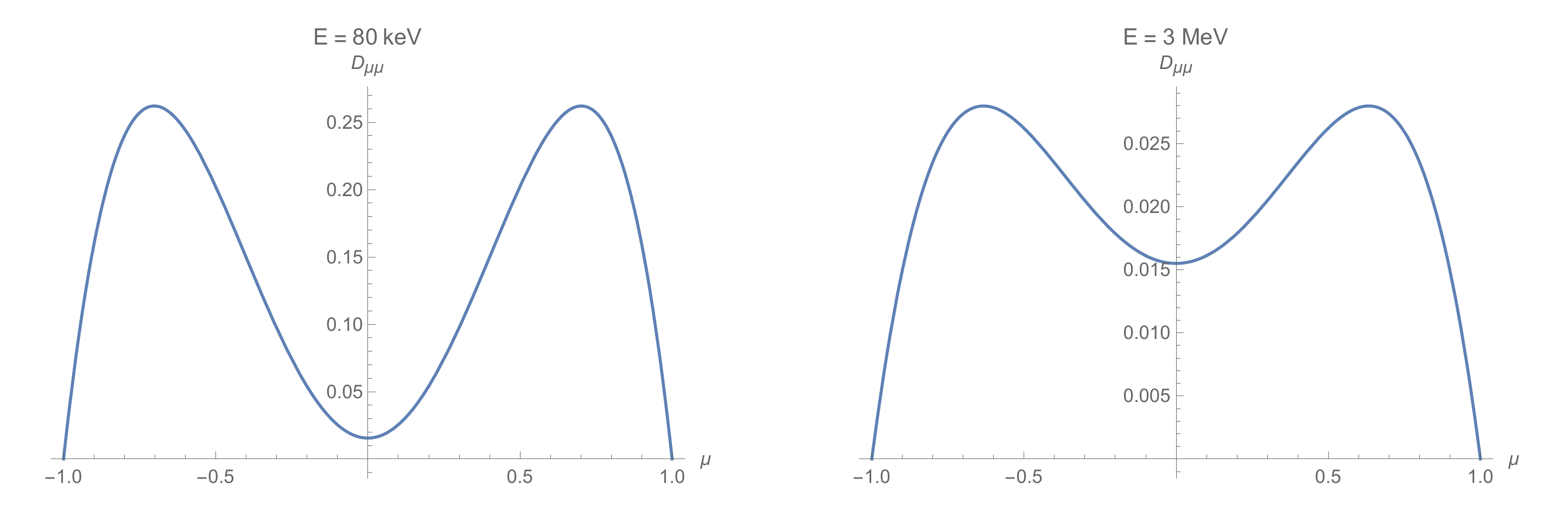}
	\end{center}
	\caption{Model calculation for electrons at extreme ends: We assume a power-law spectrum with with spectral index $s=3$ in the range $0.01 < k < M^{1/2}$. Electrons at 80 keV and 3 MeV are considered. $\Omega_p=5\times 10^4$ Hz, $v_A=0.001 c$.}
	\label{fig:modell2}
\end{figure}

A sketch of the resulting total Fokker-Planck scattering coefficient  is shown in Fig. \ref{fig:modell2}.

\section{Numerical methods}

\subsection{Particle-in-Cell simulations}

To be able to model dispersive waves of different kinds and to obtain a 
self-consistent description of electromagnetic fields and charged particles in the 
plasma, we employ a fully kinetic Particle-in-Cell (PiC) approach \cite{
hockney_1988}. In particular, we use the explicit second order PiC code \emph{
ACRONYM} \cite{kilian_2011}, which is fully relativistic, parallelized and 
three dimensional. Although the PiC method might not be the most efficient 
numerical technique when dealing with proton effects, we still favor this 
approach for its versatility. A more detailed discussion of advantages and 
drawbacks, as well as a direct comparison of PiC and MHD approaches to the 
specific problem of the interaction of protons and left-handed waves, can be 
found in  Sects. 3 and 6 of Schreiner \cite{schreiner_2014_a}. However, the PiC 
approach is well-suited for the study of electron scattering, since the time 
and length scales of electron interactions are closer to the scales of time 
step lengths and cell sizes in PiC simulations, thus reducing computing time 
compared to simulations in which proton interactions are studied.

The details of the inner workings of a PiC code shall not be discussed here. 
The simulation technique used here does not differ from standard techniques. 
Two points however are relevant for the discussion: The initialization of 
turbulence and the tracking of test particles. Turbulence will be discussed 
in Sec. \ref{sec:results_turbulence_validation_setup} as the numerical 
treatment is inherently connected to the physical processes of turbulence. 
The treatment of energetic particles is divided into to two parts: The 
injection of particles and the analysis of the particle data.

\subsubsection{Setup of turbulence simulations}
\label{sec:results_turbulence_validation_setup}

\begin{figure}[h!tb]
	\centering
	\includegraphics[width=0.6\linewidth]{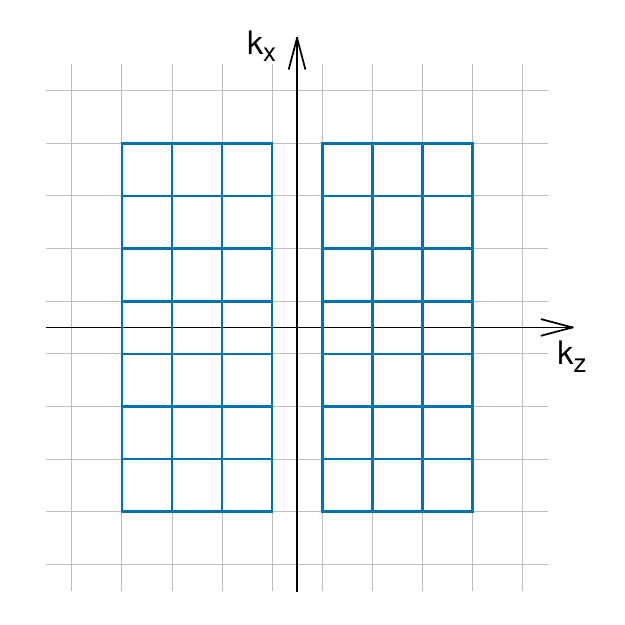}
	\caption{
	Schematic representation of two-dimensional wave number space.
	The discretized wave vectors are represented by the gray boxes, the axis 
mark the directions parallel ($k_z$) and perpendicular ($k_x$) to the 
background magnetic field $\vec{B}_0$ in the case of a two-dimensional 
simulation.
	For the simulation of decaying turbulence in two dimensions, a set of 42 
initial waves is excited, where each wave occupies one position on the grid.	
These positions are indicated by the blue boxes, in accordance with the setup 
specified by Gary\cite{gary_2008}.
	}
	\label{fig:turbulence_init}
\end{figure}

We use a simulation setup that has been inspired by Gary et al. \cite{gary_2008}: Waves are initialized at low $k$ (at $k<\Omega_i/V_A$). The layout of 
the initial waves in wave number space is explained below and sketched in Fig.
~\ref{fig:turbulence_init} for the two-dimensional setup. In the PiC simulations the velocity space and the electromagnetic fields, however, are fully three-dimensional.

As will be seen later the two-dimensional simulations show a similar energy cascade as corresponding three-dimensional simulations. Gary et al. \cite{gary_2012} highlight that the three-dimensional case differs mainly in the anisotropy and a break at $k_\perp c/\Omega_e$. The question whether particle transport is different in two and three dimensions may be answered by the fact that particle motion is still three-dimensional. The theoretical description is assuming gyrotropy anyway, the different perpendicular directions are therefore averaged out.

We employ the natural mass ratio $m_p/m_e=1836$. Waves are initialized with equal 
amplitudes and a random phase angle. The total magnetic energy density of the 
initial waves is set to 10\% of the energy density of the background magnetic 
field. This can be expressed by $\delta \!
 B^2 / B^2_0 = 0.1$, where $\delta \! B^2 = \sum_j \delta \! B^2_j$ and $j$ 
is used to index the individual waves.

To analyze the simulations, the spectra of the magnetic energy density $E_B = 
|\vec{B}^2 (\vec{k})| / (8 \, \pi)$ in wave number space are considered. A 
two-dimensional energy spectrum, i.e.~$E_B (k_\parallel, \, k_\perp)$, can be 
obtained by Fourier transforming the field data to obtain the perpendicular coordinate 
$k_\perp$. 
The parallel direction is equivalent to the z-direction of the simulation, 
whereas the perpendicular direction is represented by the x-direction in a 2D 
simulation or by the x-y-plane in a 3D simulation.
A one-dimensional spectrum $E_B (k)$ can be obtained by 
integrating over the angle $\theta$ in the $k_\parallel$-$k_\perp$-plane.
Additional one-dimensional spectra $E_B (k_\parallel)$ and $E_B (k_\perp)$ 
are obtained by integrating $E_B (k_\parallel, \, k_\perp)$ over $k_\perp$ 
and $k_\parallel$, respectively.

We study electron transport in two simulations: S1 and S2. The basic idea is 
to resolve several wave numbers in both 
the undamped and the damped regime of the whistler mode. This should allow to 
see differences in the spectral slope or the anisotropy in both regimes.
To resolve the relatively large spatial scales of the undamped regime, large 
simulation boxes are required. Thus, it was decided to restrict the 
investigations to two-dimensional setups.

Simulations S1 and S2 are characterized by the physical and numerical parameters listed in Tables~\ref{tab:simulation_turbulence_phys} and~\ref{tab:simulation_turbulence_num}.
The setups are meant to simulate decaying turbulence with a set of 42 initially excited whistler waves according to Fig. \ref{fig:turbulence_init}

\begin{table}[h]
	\centering
%	\resizebox{\columnwidth}{!}{%

	\begin{tabular}{c c c c c c}
		\hline
		\noalign{\smallskip}
		simulation & $\omega_\mathrm{p,e} \, (\mathrm{rad ~ s}^{-1})$ & $|\Omega_\mathrm{e}| \, (\omega_\mathrm{p,e})$ & $v_\mathrm{th,e} \, (c)$ & $\delta \! B^2 / B^2_0$ & $\beta$ \\
		\noalign{\smallskip}
		\hline
		\noalign{\smallskip}
		S1 & $1.966 \cdot 10^8$ & $0.447$ & $0.10$ & $0.10$ & $2.0 \cdot 10^{-1}$ \\
		\noalign{\smallskip}
		S2 & $1.966 \cdot 10^8$ & $0.447$ & $0.05$ & $0.10$ & $5.0 \cdot 10^{-2}$ \\
		\noalign{\smallskip}
		\hline
	\end{tabular}
%	}
	\caption{Physical parameters for simulations S1 and S2:
	plasma frequency $\omega_\mathrm{p,e}$, cyclotron frequency $\Omega_\mathrm{e}$, and thermal speed $v_\mathrm{th,e}$ of the electrons, the sum $\delta \! B^2$ of the squares of the magnetic field amplitudes of the individual waves, and the plasma beta $\beta$.
	%Note that the physical parameters for T6 are the same as for simulations T1 and T2, listed in Table~\ref{tab:simulation_turbulence_gary_phys}.
	}
	\label{tab:simulation_turbulence_phys}
\end{table}

\begin{table}[h]
	\centering
%	\resizebox{\columnwidth}{!}{%
	\begin{tabular}{c c c c c c c c}
		\hline
		\noalign{\smallskip}
		simulation & $N_\parallel \, (\Delta x)$ & $N_\perp \, (\Delta x)$ & $N_t \, (\Delta t)$ & $\Delta x \, (c \, \omega_\mathrm{p,e}^{-1})$ & $\Delta t \, (\omega_\mathrm{p,e}^{-1})$ & ppc\\
		\noalign{\smallskip}
		\hline
		\noalign{\smallskip}
		S1 & $2048$ & $2048$ & $1.0 \cdot 10^5$ & $7.0 \cdot 10^{-2}$ & $4.1 \cdot 10^{-2}$ & $256$\\
		\noalign{\smallskip}
		S2 & $2048$ & $2048$ & $1.0 \cdot 10^5$ & $3.5 \cdot 10^{-2}$ & $2.0 \cdot 10^{-2}$ & $256$\\
		\noalign{\smallskip}
		\hline
	\end{tabular}
%	}
	\caption{
	Numerical parameters for the two-dimensional simulations S1 and S2:
	number of cells $N_\parallel$ and $N_\perp$ in the directions parallel and perpendicular to the background magnetic field, number of time steps $N_t$, grid spacing $\Delta x$, time step length $\Delta t$, and ppc, i.e.~the number of particles (electrons and protons combined) per cell.
	}
	\label{tab:simulation_turbulence_num}
\end{table}

\subsubsection{Turbulence Spectra}
\label{sec:results_turbulence_spectra}

The simulations S1 and S2 using the parameters given above shall be discussed in short here. 

\begin{figure}[htb]
	\centering
	\includegraphics[width=0.9\linewidth]{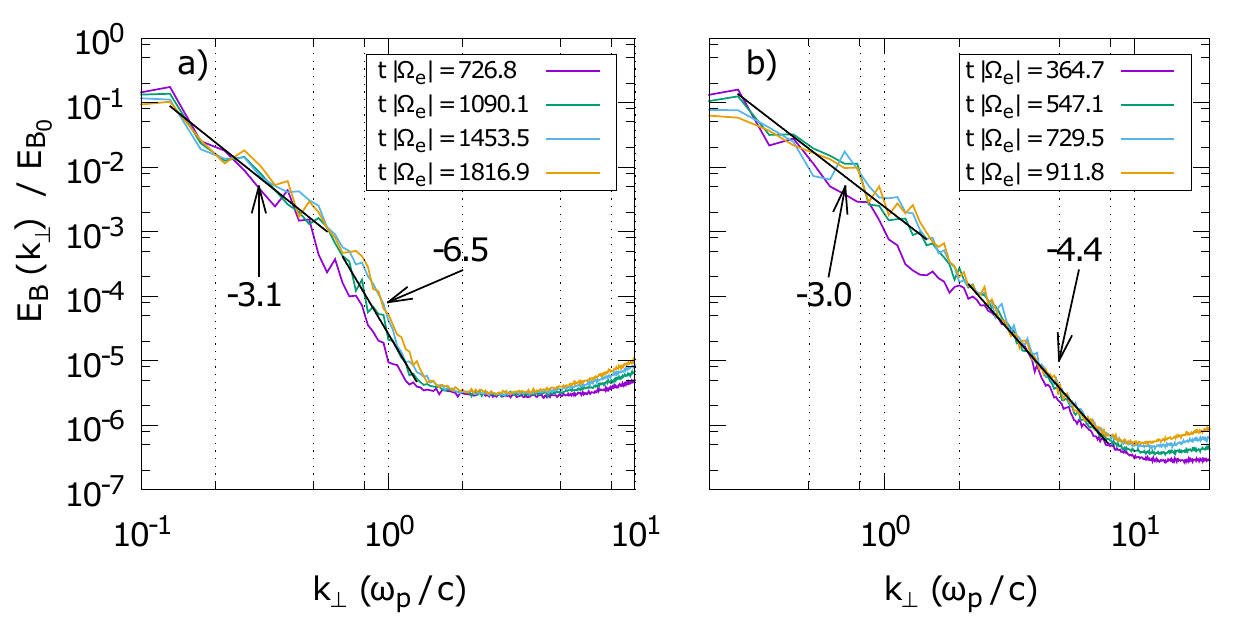}
	\caption{
	Magnetic field energy distribution $E_B (k_\perp)$ over the perpendicular wave number $k_\perp$ for simulations S1 (panel a) and S2 (panel b).
	The data is obtained at four different points in time in the two simulations, as indicated in the key.
	At the earliest time steps shown the spectra reach their steady states.
	Later in the simulations, the shapes of the spectra hardly change.
	Power law fits to the data in panels a) and b) are indicated by the black lines at times $t \, |\Omega_\mathrm{e}| = \{1090.1, \, 547.1\}$, respectively.
	The spectral indices are denoted by the numbers in the plot.
	}
	\label{fig:turbulence_production_2d_spectrum_perpendicular}
\end{figure}

Figure~\ref{fig:turbulence_production_2d_spectrum_perpendicular} shows the perpendicular spectra $E_B (k_\perp)$ of the magnetic field energy from simulations S1 (panel a) and S2 (panel b).
At small perpendicular wave numbers the magnetic energy distribution follows a power law with spectral index $\sigma_\perp = \{-3.1, \, -3.0\}$ for S1 and S2, respectively.
After the break the spectra steepen, and significant differences between both simulations become obvious in the different spectral indices.

\begin{figure}[h!tb]
	\centering
	\includegraphics[width=0.9\linewidth]{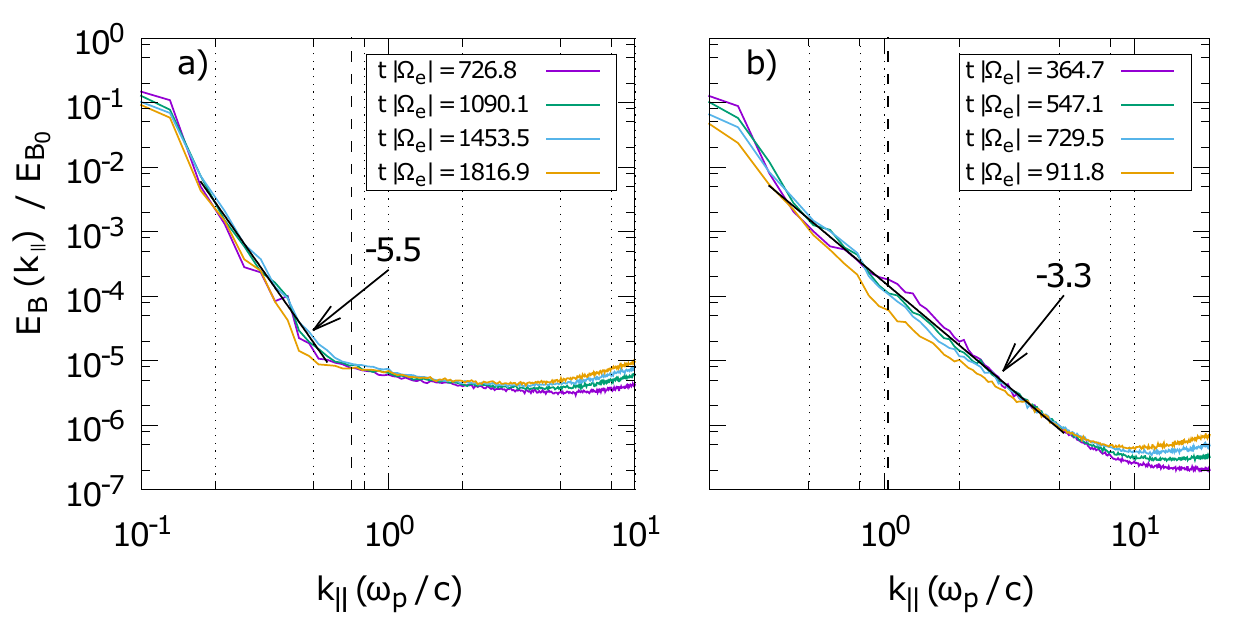}
	\caption{
	Magnetic field energy distribution $E_B (k_\parallel)$ over the parallel wave number $k_\parallel$ for simulations S1 (panel a) and S2 (panel b).
	The data is obtained at four different points in time in the two simulations, as indicated in the key.
	The spectra reach their steady state at the earliest time steps shown.
	Power law fits to the data in panels a) and b) are indicated by the black lines at times $t \, |\Omega_\mathrm{e}| = \{1090.1, \, 547.1\}$, respectively.
	The spectral indices are denoted by the numbers in the plot.
	The dashed, vertical lines mark the expected onset of cyclotron damping for purely parallel propagating waves.
	}
	\label{fig:turbulence_production_2d_spectrum_parallel}
\end{figure}

The numerical noise level in simulation S2 is about one order of magnitude lower than in S1, which allows an energy cascade to higher wave numbers.
This can be explained by the lower plasma temperature in S2, leading to less kinetic energy of the particles and therefore less fluctuations in the electromagnetic fields.
The flatter spectrum in S2 (after the break; compared to S1) agrees with results from 
Chang \cite{chang_2013}, who report a more efficient perpendicular energy transport with decreasing plasma beta $\beta$.

Chang \cite{chang_2013} also observe stronger anisotropy in simulations with 
lower $\beta$. However, this is not supported by the data from simulations S1 
and S2. The parallel spectra $E_B (k_\parallel)$ are depicted in 
Fig.~\ref{fig:turbulence_production_2d_spectrum_parallel}. In both cases the parallel 
spectra do not contain a break and are steeper than the perpendicular spectra 
at small wave numbers.
For S1 the parallel spectrum reaches the noise level approximately at the 
position where cyclotron damping is assumed to set in 
(Fig.~\ref{fig:turbulence_production_2d_spectrum_parallel} a).
Figure~\ref{fig:turbulence_production_2d_spectrum_parallel} b), however, 
shows that the parallel spectrum in simulation S2 extends to wave numbers in 
the damped regime. The slope does not change at the transition into the 
dissipation range and is 
flatter than the slope in the perpendicular spectrum at corresponding $k_\perp
$. Thus, the parallel energy transport is assumed to dominate at large wave 
numbers. Unfortunately, the turbulent cascade reaches the numerical noise 
level prior 
to that.

\begin{figure}[htb]
	\centering
	\includegraphics[width=\linewidth]{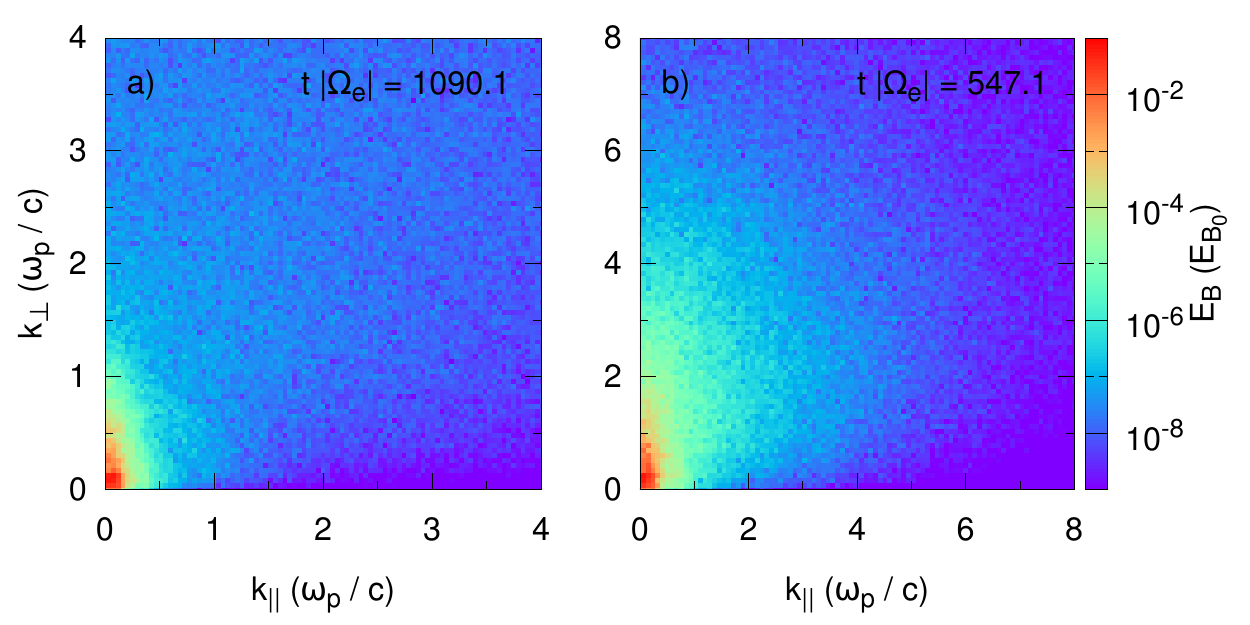}
	\caption{
	Two dimensional magnetic field energy distribution in wave number space for simulations S1 (panel a) and S2 (panel b).
	Note the different scales on the axis of both panels.
	}
	\label{fig:turbulence_production_2d}
\end{figure}

The energy distribution $E_B (k_\parallel, \, k_\perp)$ in two-dimensional wave number space supports the claim that parallel energy transport becomes important in simulation S2, as Fig.~\ref{fig:turbulence_production_2d} shows.
Panel b) depicts the distribution of magnetic field energy in simulation S2.
Although hardly any (quasi-) parallel waves are produced above $k_\parallel \, c / \omega_\mathrm{p} \approx 1$ (where cyclotron damping sets in), this critical parallel wave number can be passed at higher $k_\perp$.
At small wave numbers, however, the perpendicular cascade clearly dominates.
In simulation S1 the situation is different, as Fig.~\ref{fig:turbulence_production_2d} a) shows.
The perpendicular cascade at small wave numbers is similar to S2, as expected, but at larger $k_\perp$ there is hardly any energy transport to higher parallel wave numbers, in agreement with the spectra in Figs.~\ref{fig:turbulence_production_2d_spectrum_perpendicular} and~\ref{fig:turbulence_production_2d_spectrum_parallel}.

\subsection{Simulation of energetic particles}

\label{sec:numerics_test_particles}

In order to study wave-particle scattering, a specific initialization of a 
test particle population is prepared. The ACRONYM code allows for different 
particle species (typically protons and electrons, but also positrons or 
heavier ions) and different particle populations (a background plasma and, 
for example, additional jet populations, non-thermal particles, etc.).

The simulations S1 and S2 discussed here employ a thermal background plasma 
(see Table \ref{tab:simulation_turbulence_phys}) and an additional population of non-thermal test particles to study the transport of energetic electrons. By doing so the test particles have no noticeable influence 
on the background plasma, even if the ratio of numerical particles $N_\mathrm{
t} / N_\mathrm{bg}$ in the test and the background particle population is of 
the order of unity.

\subsubsection{Initialization and analysis}
\label{sec:numerics_test_particles_init}

Test particles are initialized as a mono-energetic population, i.e.~the 
particles have the same absolute speed, but their direction of motion is 
chosen randomly. The speed is calculated from the resonance condition for 
waves in the plasma. Solving Eq.~(\ref{eq:resonance_condition}) for the speed 
of a particle of species $\alpha$ yields:
\begin{equation}
	v_\alpha = \left|\frac{k_\parallel \, \omega \, |\mu_\mathrm{res}| \pm |
\Omega_\alpha| \, \sqrt{k_\parallel^2 \, \mu_\mathrm{res}^{2} + (\Omega_\alpha
^2 - \omega^2) / c^2}}{k_\parallel^2\,\mu_\mathrm{res}^{2} + \Omega_\alpha^2 /
 c^2}\right|,
	\label{eq:test_particle_speed}
\end{equation}
where $\mu_\mathrm{res}$ is the desired resonant pitch angle cosine.
The sign in the numerator changes depending on the polarization of the wave, 
its direction of propagation, and the particle species.

The directions of motion of the bulk of the test particles are chosen at 
random, using the speed calculated from Eq.~(\ref{eq:test_particle_speed}), a 
random polar angle cosine $\mu$, and a random azimuth angle $\phi$.
This yields an isotropic distribution of the velocity vectors in $\mu$-$\phi$-
space. It is convenient to choose an isotropic distribution in $\mu = \cos 
\theta$ (instead of $\theta$), because the analysis of pitch angle scattering 
relies on the pitch angle cosine and not on the pitch angle itself.

\begin{figure}[htb]
	\centering
	\includegraphics[width=\linewidth]{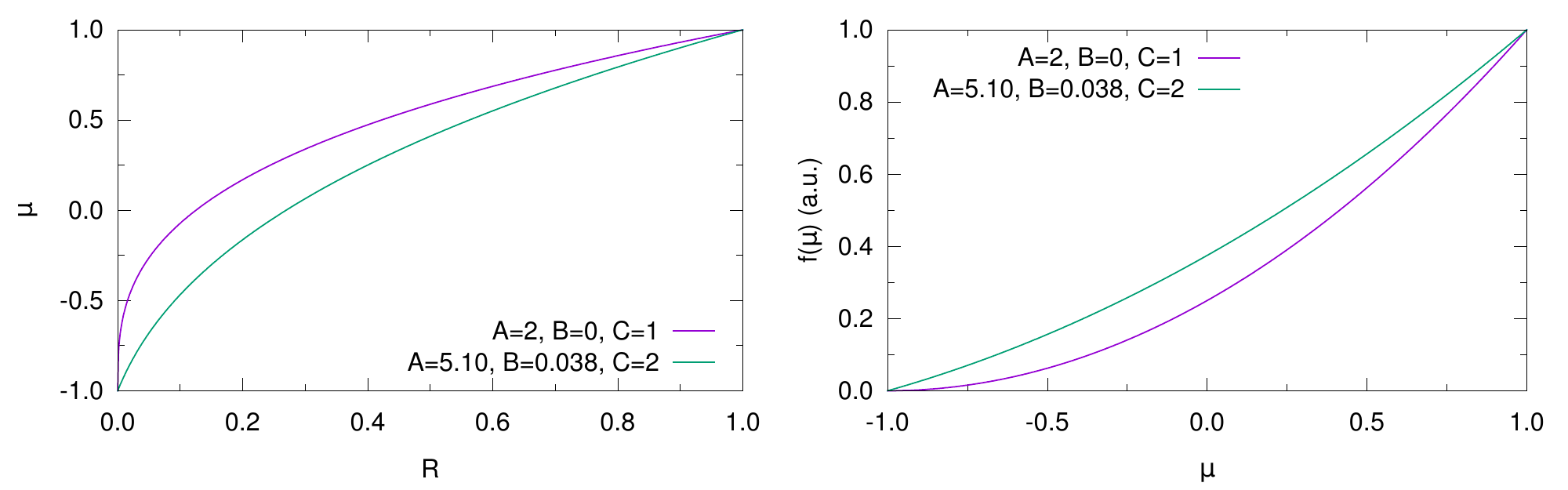}
	\caption{
	A fraction of the test particle population has their pitch angle cosines 
assigned according to a parabolic distribution.	The left panel shows the 
assigned pitch angle cosine $\mu$ as a function of the random number $R \in [0
,1]$, which is used to generate the distribution.	The curves in the left 
panel follow Eq.~(\ref{eq:pitch_angle_parabola}) and employ two sets of 
parameters $A$, $B$, and $C$, as indicated.	The right panel depicts the 
resulting particle distribution $f(\mu)$ as a function of $\mu$.
	The purple lines employ the parameters suggested by Ivascenko et al. \cite{
ivascenko_2016}, whereas the green curves show an improved implementation which 
is used in the ACRONYM code.
	Note that the derivative $d f(\mu) / d\mu \neq 0$ over the whole range of 
pitch angle cosines in the latter case, whereas it becomes zero at $\mu = -1$ 
when the parameters of Ivascenko et al. \cite{ivascenko_2016} are used.
	}
	\label{fig:pitch_angle_parabola}
\end{figure}

A fraction of the test particle population is not initialized as described 
above, but instead uses a parabolic distribution of polar angle cosines.
This is done by assigning
\begin{equation}
	\mu = A \, (R+B)^{1/3} - C
	\label{eq:pitch_angle_parabola}
\end{equation}
to the particles, where $R$ is a random number between zero and one and $A$, $
B$, and $C$ are parameters describing the shape of the parabola.
The parabolic distribution is required for the analysis of pitch angle scattering using the method of Ivascenko et al.\cite{ivascenko_2016}. Ivascenko et al. suggest the use of a half-parabola, i.e.~$A=2$, $B=0$, $C=1$, but other distributions are also possible  (see Fig.~\ref{fig:pitch_angle_parabola}). The resulting angular distribution of the entire test particle 
population is, therefore, not entirely isotropic.

The technique described above to create a population of energetic test 
particles for the study of wave-particle scattering 
was designed for a single plasma wave in the simulation 
\cite{schreiner_2017_b}.  However, it can also be applied to simulations with several plasma waves. 

The test particle population is not injected at the start of the simulations 
S1 and S2, but at a later time $t_\text{inj}$ for the following reason: It is 
expected that turbulence develops from 
the initial conditions of the simulation, i.e.~from a small set of seed waves 
which interact and start the turbulent cascade. This process takes time and 
it may be desired to wait until a turbulent cascade is established before the 
transport of energetic test particles can be studied. Therefore an optional 
deployment of test particles at later times in the simulation is favored. The 
particles are created at a a pre-defined time step and the initialization  is 
carried out as described earlier. They can then be tracked for the rest of the 
simulation.

To evaluate particle transport in the turbulent plasma the test particle data can be analyzed after the simulation to obtain the diffusion coefficient $D_{\mu\mu}$. The diffusion coefficient is calculated from a simplified Fokker-Planck equation, Eq.~(\ref{eq:fp}), where pitch angle diffusion is assumed to be the only relevant diffusion process:
\begin{equation}
	\frac{\partial f_\alpha}{\partial t} - \frac{\partial}{\partial \mu} \, D_{\mu\mu} \, \frac{\partial f_\alpha}{\partial \mu} = 0.
	\label{eq:fokker_planck_equation_simplified}
\end{equation}
This equation can be rewritten to yield
\begin{equation}
	\frac{\partial f_\alpha (\mu,t)}{\partial t} = \left( \frac{d D_{\mu\mu} (\mu)}{d \mu} \right) \, \frac{\partial f_\alpha (\mu,t)}{\partial \mu} + D_{\mu\mu} (\mu) \, \frac{\partial^2 f_\alpha (\mu,t)}{\partial \mu^2}.
	\label{eq:fokker_planck_equation_simplified_derivative}
\end{equation}
The  method described by Ivascenko et al. \cite{ivascenko_2016} is based on integrating Eq.~(\ref{eq:fokker_planck_equation_simplified}) over $\mu$, which yields the pitch angle current $j_\mu$:
\begin{equation}
	\int\limits^\mu_{-1} \mu \, \frac{\partial f_\alpha (\mu,t)}{\partial t} = D_{\mu\mu} (\mu) \, \frac{\partial f_\alpha (\mu,t)}{\partial \mu} = - j_\mu.
	\label{eq:fokker_planck_equation_simplified_integral}
\end{equation}
The diffusion coefficient is then obtained by dividing $j_\mu$ by $\partial f_\alpha / \partial \mu$.

\subsubsection{Physical parameters}

Using the setups of simulations S1 and S2, the transport of energetic 
electrons in kinetic turbulence is studied. In the following, the exact 
parameters for the test particle energy distribution are presented

In simulations S1 and S2 decaying whistler turbulence is simulated, as was 
shown in the Section \ref{sec:numerics_test_particles}. As can be seen in the 
magnetic energy spectra 
presented in Figs.~\ref{fig:turbulence_production_2d_spectrum_perpendicular} 
and~\ref{fig:turbulence_production_2d_spectrum_parallel}, a steady state in 
terms of the power law slope of the spectral energy distribution is 
established after a given time in each of the two simulations
%\footnote{
%Since decaying turbulence is simulated, the total energy decreases with time.
%The ``steady state'' therefore does not imply that the spectral energy distribution is constant over time, but rather that the shape of the distribution does not change qualitatively.
%}.
As soon as this stage of the simulation is reached, a population of energetic test electrons can be injected as described in Sec. \ref{sec:numerics_test_particles_init}.

The time step for the checkpoint and subsequent restart is chosen to be $t \, 
|\Omega_\mathrm{e}| = 726.8$ for S1 and $t \, |\Omega_\mathrm{e}| = 364.7$ 
for S2.
%Simulations T6 and T7 represent the setups of the background plasma, as 
%defined in Tables~\ref{tab:simulation_turbulence_phys} and~\ref{tab:simulation
%_turbulence_num}.
For each of these two setups six test electron configurations are prepared.
The simulations are labeled according to the physical setup (S1 or S2) 
followed by a letter referring to the test particle configuration (``a'' 
through ``f''). The parameters of the test particles can be found in Table~
\ref{tab:simulation_turbulence_particles} and describe the test electron 
speed $v_\mathrm{e}$ and kinetic energy $E_\mathrm{kin,e}$.

\begin{table}[h]
	\centering
	\begin{tabular}{c c c c c c c}
		\hline
		\noalign{\smallskip}
		simulation & S$j$a & S$j$b & S$j$c & S$j$d & S$j$e & S$j$f \\
		\noalign{\smallskip}
		\hline
		\noalign{\smallskip}
		$v_\mathrm{e} \, (c)$ & $0.546$ & $0.862$ & $0.941$ & $0.979$ & $0.999$ & $0.999$ \\
		\noalign{\smallskip}
		$E_\mathrm{kin,e} \, (\mathrm{eV})$ & $1.0 \cdot 10^5$ & $5.0 \cdot 10^5$ & $1.0 \cdot 10^6$ & $2.0 \cdot 10^6$ & $1.0 \cdot 10^7$ & $1.0 \cdot 10^7$ \\
		\noalign{\smallskip}
		\hline
	\end{tabular}
	\caption{
	Test electron characteristics for the simulations of particle transport:
	test electron speed $v_\mathrm{e}$ and corresponding kinetic energy $E_\mathrm{kin,e}$.
	The individual simulations (letters ``a'' through ``f'') are based on the simulations of kinetic turbulence S$j$, with $j \in \{1, \, 2\}$, which are described in Sect.~\ref{sec:results_turbulence_spectra} (Tables~\ref{tab:simulation_turbulence_phys} and~\ref{tab:simulation_turbulence_num}).
	Note that simulations S$j$e and S$j$f employ the same test electron energies.
	However, they differ in the way the test electron distribution is initialized (see text).
	}
	\label{tab:simulation_turbulence_particles}
\end{table}

The test electron energy is increased from simulation S1a (S2a) to S1e (S2e).
Simulation S1f (S2f) uses the same particle energy as S1e (S2e), but a different parabolic angular distribution of the particles:
Here the particle density $f(\mu)$ increases with increasing $\mu$, while in the other simulations it decreases with increasing pitch angle cosine.
This change in the pitch angle distribution allows to check for systematic errors in the particle data.

\section{Results}

\subsection{Pitch-angle diffusion coefficients}
\label{sec:results_turbulence_transport_diffusion}

\begin{figure}[htp]
	\centering
	\includegraphics[width=\linewidth]{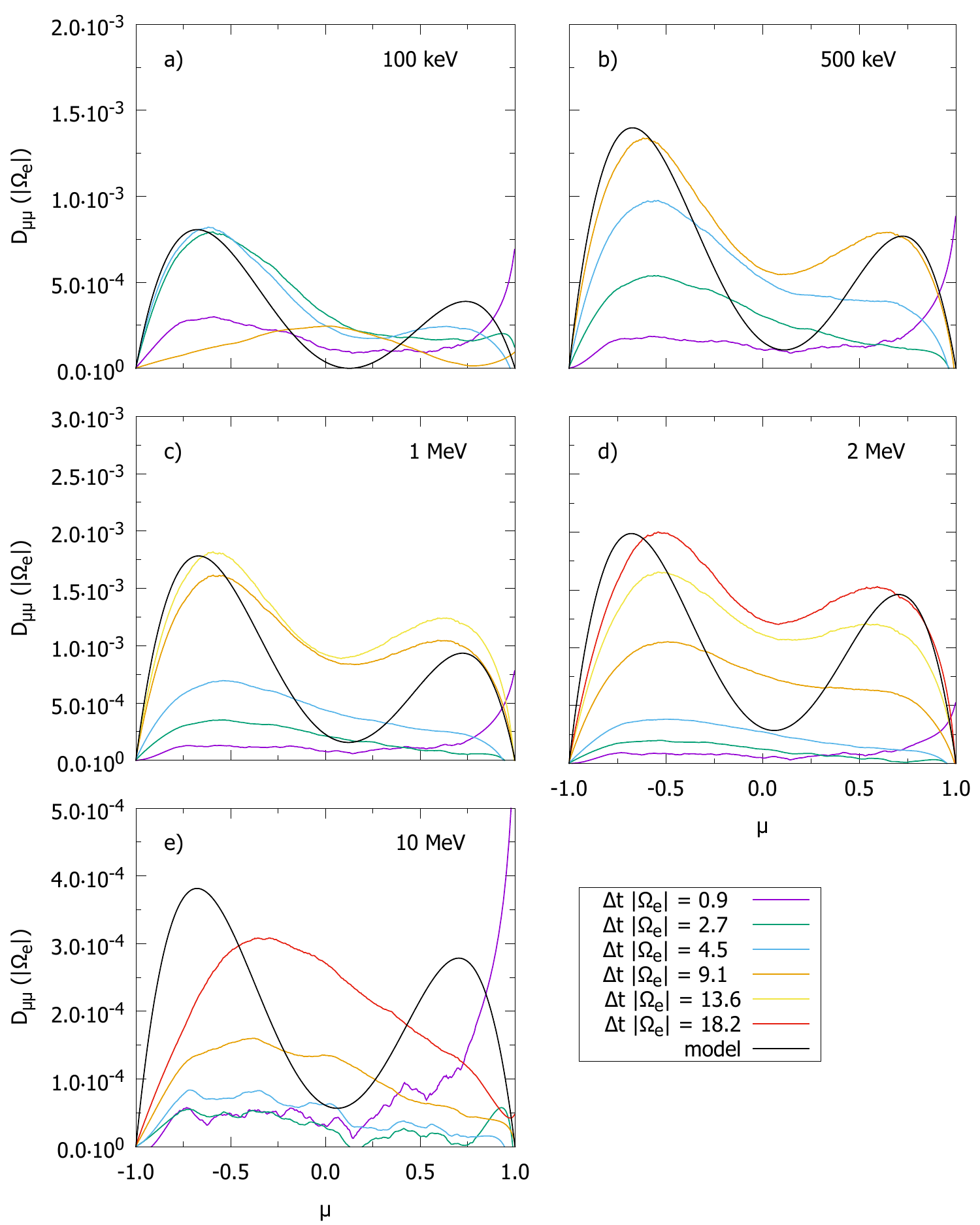}
	\caption{
	Pitch angle diffusion coefficients $D_{\mu\mu}$ for test electrons with different energies as indicated in the individual panels.
	Panels a) through e) refer to simulations S1a through S1e.
	The colored lines denote the diffusion coefficients derived from the simulation data at various times.
	The black lines follow the model predictions derived in Sec. \ref{sec:transport:electron}.
	}
	\label{fig:particles_Dmumu_set10}
\end{figure}

\begin{figure}[htp]
	\centering
	\includegraphics[width=\linewidth]{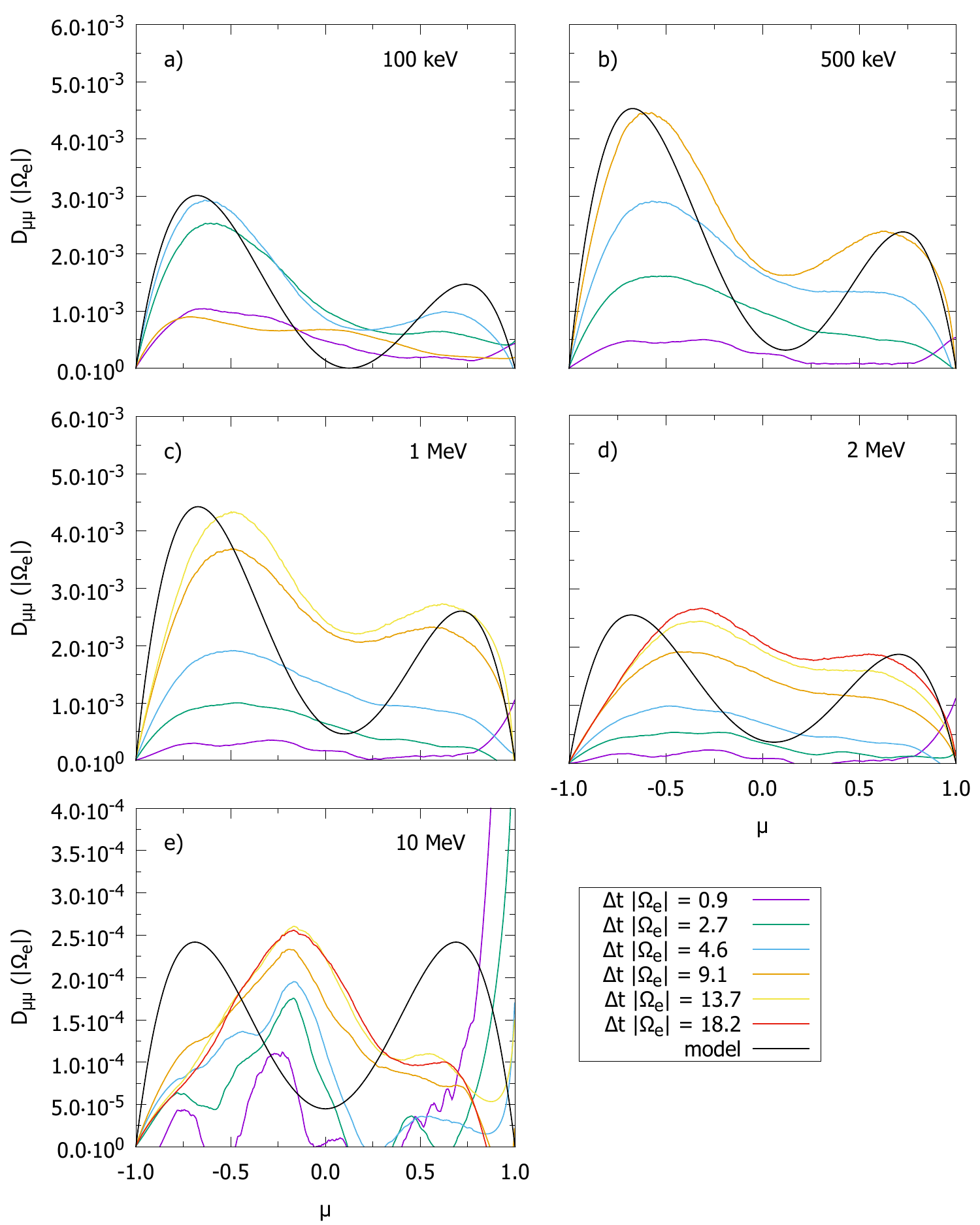}
	\caption{
	Pitch angle diffusion coefficients $D_{\mu\mu}$ for test electrons with different energies as indicated in the individual panels.
	Panels a) through e) refer to simulations S2a through S2e.
	The colored lines denote the diffusion coefficients derived from the simulation data at various times.
	The black lines follow the model predictions derived in Sec. \ref{sec:transport:electron}.
	}
	\label{fig:particles_Dmumu_set11}
\end{figure}

The test particle simulations are analyzed as described 
in Sec. \ref{sec:numerics_test_particles_init}.
The energetic electrons are tracked for several electron cyclotron time 
scales and the resulting pitch angle diffusion coefficients $D_{\mu\mu}$ are 
presented in Figs.~\ref{fig:particles_Dmumu_set10} and~\ref{fig:particles_Dmumu_set11} for data based on the setup of S1 and S2, respectively. 
Time is measured as the interval $\Delta t$ from the time of the injection of 
the particles to the current time step.

The results of both sets of simulations, one based on S1, the other based on S2, 
do not differ qualitatively, as would be expected from the two setups. The 
only difference between the physical parameters for S1 and S2 is the plasma 
temperature, which has no direct influence on the test electrons. Although 
the magnetic energy spectrum differs at high wave numbers (see 
Fig.~\ref{fig:turbulence_production_2d_spectrum_perpendicular}), the 
distribution of magnetic energy at small wave numbers is very similar. As the 
explanations below will show, this low wave number regime represents the dominant 
influence on particle transport. Thus, the two sets of simulations will be 
discussed simultaneously in the following.

Although particle data can be obtained for an arbitrary number of time steps, 
the interval which can be used for the analysis is still limited. The method 
of Ivascenko \cite{ivascenko_2016} critically depends on the particle 
distribution $f(\mu)$ in pitch angle space. In order for the method to work, 
the initial distribution must be slightly disturbed, but the perturbations 
must not be too strong. This leaves only a brief period of time for the 
optimal efficiency of the method.

Figures~\ref{fig:particles_Dmumu_set10} a) and \ref{fig:particles_Dmumu_set11}
 a) show the typical behavior of the derived $D_{\mu\mu}$ over time. Shortly 
after the injection of the test electrons the perturbations of $f (\mu)$ are 
small, resulting in a low amplitude of $D_{\mu\mu}$ (purple lines). With 
increasing time the amplitude grows and reaches a maximum (green and blue 
lines). At later times the amplitude decreases again, as the perturbations 
become too strong and the method becomes unreliable (orange lines). The other 
panels of the two figures show the time evolution until the maximum amplitude 
of $D_{\mu\mu}$ is reached for other test electron energies.

Although the turbulent cascade is assumed to be symmetric about $\mu = 0$, 
the panels of Figs.~\ref{fig:particles_Dmumu_set10} 
and~\ref{fig:particles_Dmumu_set11} show an obvious asymmetry in the pitch
angle diffusion coefficients derived from the test electron data.
The amplitude of $D_{\mu\mu}$ is generally larger for $\mu < 0$.
While the energy spectrum itself is isotropic in $\mu$, one could argue that 
the polarization of the waves' magnetic fields relative to the direction of 
the background magnetic field $B_0$ is different (i.e.~the plasma physics 
definition of the polarization).

%This should, however, not have an influence on the scattering behavior of the 
%test electrons. Despite the opposite polarization of the waves (in the plasma physics notation
%) the resulting scattering amplitudes are symmetric about $\mu = 0$.

\begin{figure}[htb]
	\centering
	\includegraphics[width=\linewidth]{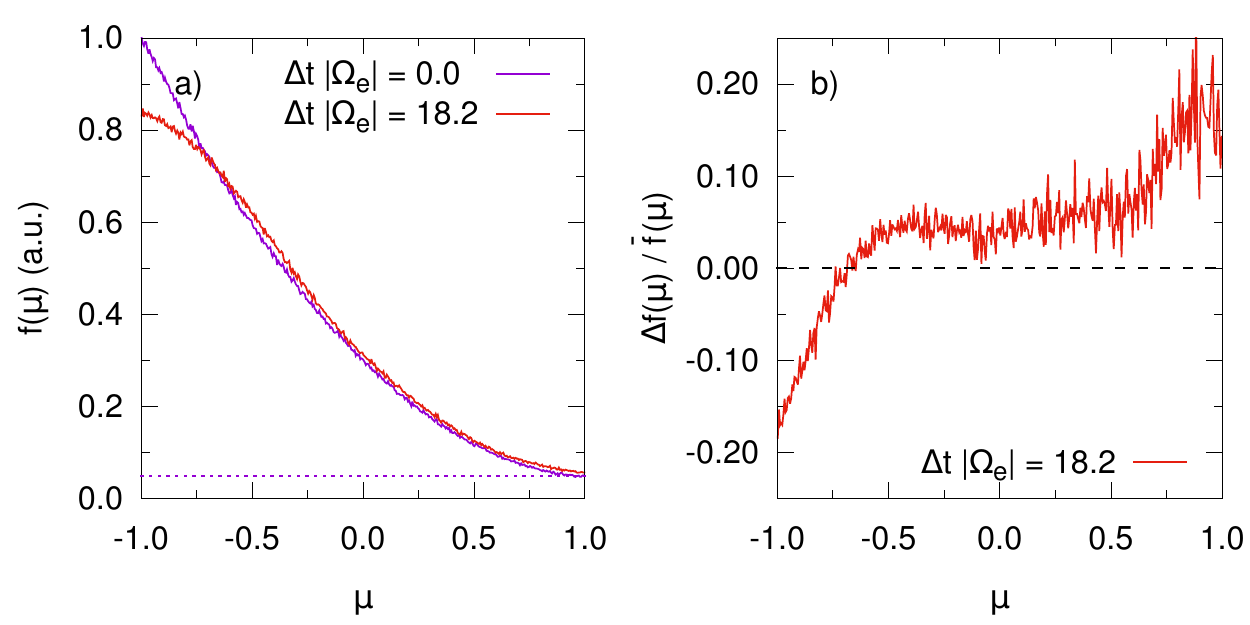}
	\caption{
	Test electron distribution in pitch angle space.
	Panel a) shows the initial distribution $f(\mu)$ at the time of the injection of the test electrons ($\Delta t = 0$) and at a later time in simulation S1c.
	The relative deviation $\Delta f / \bar{f}$ of the distributions at these two time steps is depicted in panel b).
	The deviation is defined as the difference of the two distributions over their mean value.
	It can be seen that statistical fluctuations become more significant for larger $\mu$.
	}
	\label{fig:particles_parabola_histo}
\end{figure}

The magnetic helicity of the plasma waves is one of the possible causes of this anisotropy. 
Another reason for the asymmetry found in 
Figs.~\ref{fig:particles_Dmumu_set10} and~\ref{fig:particles_Dmumu_set11} is that the parabolic distribution $f(
\mu)$ of the test particles implies that there are more test electrons at 
negative pitch angle cosines (except for simulations S1f and S2f).
Therefore, the particle statistics is more reliable for negative $\mu$ and 
the method of Ivascenko et al.\cite{ivascenko_2016} produces more accurate 
diffusion coefficients.
While $D_{\mu\mu}$ can also be calculated for $\mu > 0$, it is more prone to 
errors and statistical fluctuations play a more important role, as 
Fig.~\ref{fig:particles_parabola_histo} indicates.
However, small scale statistical fluctuations can be suppressed by use of a 
Savitzky-Golay-filter, as suggested by Ivascenko et al. \cite{ivascenko_2016}.

Especially for early time steps it can be seen that $D_{\mu\mu}$ is found to 
diverge at $\mu = 1$. This is, of course, not a physical effect.
At $\mu = 1$ the derivative of the initial parabolic distribution $f(\mu)$ 
becomes (almost) zero. In this case, the method of Ivascenko et al. 
\cite{ivascenko_2016} becomes numerically unstable.

Another numerical effect causes $D_{\mu\mu}$ to become negative.
This can be seen in panels d and e of Fig.~\ref{fig:particles_Dmumu_set11} 
for early times. Negative solutions are most likely related to statistical 
fluctuations in the particle distribution, which drown the signal at early 
times, when the physically motivated perturbations of $f(\mu)$ are still 
developing.

Besides these flaws, the derived pitch angle diffusion coefficients appear 
reasonable. They develop a (more or less) symmetric shape about $\mu = 0$, 
indicating that neither direction is preferred. This is expected from the 
setup of the turbulence simulations S1 and S2, which employ a symmetric 
layout of initial waves and therefore should produce turbulent cascades which 
are symmetric in $\mu$\footnote{This, however, cannot be 
proven by the plots of the energy distribution in wave number space, since 
the information about the direction of propagation of the waves is lost.
}.

An interesting observation is that the pitch angle diffusion coefficients 
grow in amplitude with the particle energy increasing from $100 \, \mathrm{keV
}$ to $2 \, \mathrm{MeV}$. At the highest test electron energy, however, the 
amplitude of $D_{\mu\mu}$ is significantly lower than in all other cases.
Both Figs.~\ref{fig:particles_Dmumu_set10} e) 
and~\ref{fig:particles_Dmumu_set11} e) also show that $D_{\mu\mu}$ forms a single peak close to $\mu = 0$ in 
the case of the highest electron energy, whereas all other simulations 
produce a double peak structure. The reason for these differences is not clear.
However, it is assumed that the different behavior of the $10 \, \mathrm{MeV}$-electrons is related to their scattering characteristics:
These high energy particles resonate with all of the initially excited waves 
in the simulations (see Fig. \ref{fig:turbulence_init}), which is not the case in the simulations of less 
energetic electrons. Since the initial waves contain the most energy, they 
are also assumed to significantly influence particle transport, especially if 
wave-particle resonances may occur.

\begin{figure}[htb]
	\centering
	\includegraphics[width=0.95\linewidth]{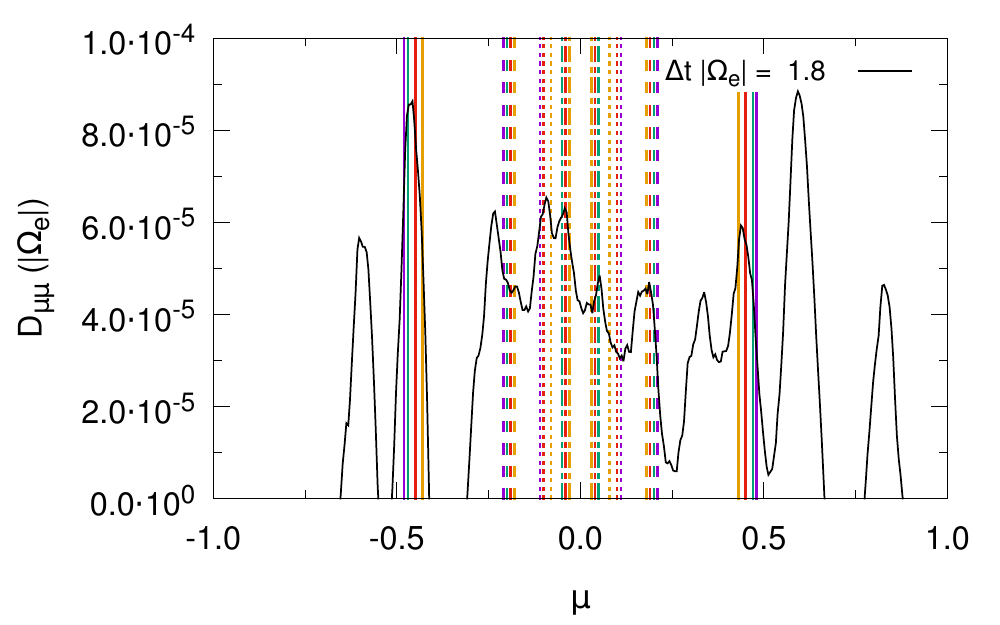}
	\caption{
	Pitch angle diffusion coefficient $D_{\mu\mu}$ at one point in time as derived from the data of simulation S1f (black line).
	A noticeable number of peaks in $D_{\mu\mu}$ coincide with the positions of wave-particle resonances predicted by the resonance condition~(\ref{eq:resonance_condition}), which are marked by the colored, vertical lines.
	The colors denote the parallel wave numbers $k_\parallel$ (in numerical units) from one to four: purple, green, red, orange.
	The line style refers to perpendicular wave numbers $k_\perp$ (also in numerical units) from zero to three: solid, dashed, dotted, dot-dashed.
	For example, the red dashed lines represent the resonance with a wave at $(k_\parallel = \pm 3, \, k_\perp = 1)$.
	Only resonances of the first order, i.e.~$n = \pm 1$ in Eq.~(\ref{eq:resonance_condition}) are shown.
	Note that $k_\perp$ does not enter the resonance condition explicitly, but is required to calculate the frequency $\omega (k_\parallel, \, k_\perp)$ according to the cold plasma dispersion relation.
	}
	\label{fig:particles_resonance}
\end{figure}

In fact, the $D_{\mu\mu}$ in Figs.~\ref{fig:particles_Dmumu_set10} e) and~\ref
{fig:particles_Dmumu_set11} e) exhibit distinct peaks at early times (purple 
and green curves). Similar behavior is also found in simulations S1f and S2f, 
which are not included in Figs.~\ref{fig:particles_Dmumu_set10} 
and~\ref{fig:particles_Dmumu_set11}. For the example of one time step in 
simulation S1f, the peak 
structures in $D_{\mu\mu}$ are related to wave-particle resonances calculated 
according to Eq.~(\ref{eq:resonance_condition}).
The result is shown in Fig.~\ref{fig:particles_resonance}, where the colored 
vertical lines mark the expected positions of resonances.
It can be seen that the resonances coincide with the positions of the peaks 
in $D_{\mu\mu}$.
The region around $\mu=0$ is most densely populated by resonances, which 
might explain the single peak in $D_{\mu\mu}$ at later times as seen in Figs.~
\ref{fig:particles_Dmumu_set10} e) and~\ref{fig:particles_Dmumu_set11} e).

Finally, Figs.~\ref{fig:particles_Dmumu_set10} and~\ref{fig:particles_Dmumu_set11} also include the model predictions from Eqs. \eqref{eq:dmmtotal1} and 	\eqref{eq:dmmtotal2}.
Some of the parameters required can be directly obtained from the setup of the simulations:
The ratio $\delta \! B^2 / B^2_0$ is listed in Table~\ref{tab:simulation_turbulence_phys}, the test electron speed speed $v_e$ and the electron cyclotron frequency $\Omega_e$ are found in Table ~\ref{tab:simulation_turbulence_particles}. However, the minimum wave number $k_\mathrm{min}$, the spectral index $s$, and the cross helicity and magnetic helicity are not as trivial to find.

For the minimum wave number the magnetic energy spectra in 
Figs.~\ref{fig:turbulence_production_2d_spectrum_perpendicular} 
and~\ref{fig:turbulence_production_2d_spectrum_parallel} have been considered.
The second smallest resolved wave number $k_\mathrm{min} = 2 \, \Delta k$ has 
been chosen, where $\Delta k$ is the grid spacing in wave number space.
In a square simulation box, where the numbers of grid cells $N_\parallel$ and 
$N_\perp$ in the parallel and perpendicular directions are equal, the grid 
spacing is given by $\Delta k = 2 \, \pi / (N_\parallel \, \Delta x) = 2 \, 
\pi / (N_\perp \, \Delta x)$.
The minimum wave number $k_\mathrm{min}$ marks the beginning of the downward 
slope of the energy spectrum.
Waves at small wave numbers are assumed to dominate the interaction with the 
particles due to their high energy content and the steep spectral slope.
Therefore, the spectral index $s = |\sigma_\perp| = 3.1$ has been chosen, in 
accordance with the index of the perpendicular spectrum in 
Fig.~\ref{fig:turbulence_production_2d_spectrum_perpendicular}~a).
This spectral index corresponds to simulation S1, but since the index in S2 
is similar, $s = 3.1$ was used in both cases.

Finally the magnetic helicity $\sigma$ was chosen to be 0. The effect is in 
fact rather small since electrons mostly resonate with right-handed polarized 
modes.

From this starting point the three parameters $k_\mathrm{min}$, $s$, and $H_R$
 were fitted according to the numerical data from each simulation.
The resulting parameters which are used in the plots in 
Figs.~\ref{fig:particles_Dmumu_set10} and~\ref{fig:particles_Dmumu_set11} are 
listed in Table~\ref{tab:model_turbulence_parameters}.
It can be seen that most simulations can be described with the initial 
choices for $k_\mathrm{min}$ and $s$ explained above.

%Only for S1a and S2a the spectral index had to be changed slightly.
%For simulations S1e, S1f, S2e, and S2f, i.e.~the simulations with the highest 
%test electron energies, the minimum wave number had to be adapted.
%However, by looking at panels e) of Figs.~\ref{fig:particles_Dmumu_set10} and~
%\ref{fig:particles_Dmumu_set11} it becomes obvious that the model by \cite{
%shalchi_2006} fails to describe the numerical data anyway, since it produces 
%two peaks, whereas the particle data suggests only one peak.

\begin{table}[h]
	\centering
% 	\begin{tabular}{c c c c c c c}
% 		\hline
% 		\noalign{\smallskip}
% 		simulation & T6a & T6b & T6c & T6d & T6e & T6f \\
% 		\noalign{\smallskip}
% 		\hline
% 		\noalign{\smallskip}
% 		$s$ & $3.0$ & $3.1$ & $3.1$ & $3.1$ & $3.1$ & $3.1$ \\
% 		\noalign{\smallskip}
% 		$\Lambda$ & $0.80$ & $0.64$ & $0.57$ & $0.48$ & $0.25$ & $0.25$ \\
% 		\noalign{\smallskip}
% 		$k_\mathrm{min} \, (\Delta k)$ & $2$ & $2$ & $2$ & $2$ & $1$ & $1$ \\
% 		\noalign{\smallskip}
% 		\hline
% 		\noalign{\smallskip}
% 		simulation & T7a & T7b & T7c & T7d & T7e & T7f \\
% 		\noalign{\smallskip}
% 		\hline
% 		\noalign{\smallskip}
% 		$s$ & $3.0$ & $3.1$ & $3.1$ & $3.1$ & $2.5$ & $2.5$ \\
% 		\noalign{\smallskip}
% 		$\Lambda$ & $0.80$ & $0.58$ & $0.47$ & $0.33$ & $0.15$ & $0.15$ \\
% 		\noalign{\smallskip}
% 		$k_\mathrm{min} \, (\Delta k)$ & $2$ & $2$ & $2$ & $2$ & $1$ & $1$ \\
% 		\noalign{\smallskip}
% 		\hline
% 	\end{tabular}
	\resizebox{\columnwidth}{!}{%
	\begin{tabular}{c c c c c c c c c c c c c c c}
		\hline
		\noalign{\smallskip}
		simulation                     & & S1a & S1b & S1c & S1d & S1e & S1f & & S2a & S2b & S2c & S2d & S2e & S2f \\
		\noalign{\smallskip}
		\hline
		\noalign{\smallskip}
		$s$                            & & $3.1$ & $3.1$ & $3.1$ & $3.1$ & $3.1$ & $3.1$ & & $3.1$ & $3.1$ & $3.1$ & $3.1$ & $3.1$ & $3.1$ \\
		\noalign{\smallskip}
		$H_R$                      & & $1.00$ & $0.55$ & $0.55$ & $0.26$ & $0.26$ & $0.25$ & & $0.99$ & $0.59$ & $0.42$ & $0.26$ & $0$ & $0$ \\
		\noalign{\smallskip}
		$k_\mathrm{min} \, (\Delta k)$ & & $2$ & $2$ & $2$ & $2$ & $1$ & $1$ & & $2$ & $2$ & $2$ & $2$ & $1$ & $1$ \\
		\noalign{\smallskip}
		\hline
	\end{tabular}
	}
	\caption{
	Parameters assumed for the model:
	spectral index $s$, cross-helicity $H_R$, and minimum wave number $k_\mathrm{min}$.
	The latter is given in units of the grid spacing $\Delta k = \{4.4, \, 8.7\} \cdot 10^{-2} \, \omega_\mathrm{p} / c$ in wave number space in simulations S1 and S2, respectively.
	}
	\label{tab:model_turbulence_parameters}
\end{table}

In general, the model describes the data surprisingly well. Position and amplitude of the maxima and the inclination of the flanks are in good agreement. The contribution at $\mu=0$ are in disagreement - this is however not unexpected as for quasi-linear theory. Still we find a non-zero contribution at medium energies, which is different from a non-dispersive QLT approach.
%keeping in mind that \cite{shalchi_2006} derived their equations for Alfv\'enic turbulence.
%The pitch angle diffusion coefficients derived from the particle data mainly exhibit a double peak structure, which is symmetric about $\mu = 0$.
%Although the model is not able to recover the shape of these two peaks exactly, their amplitude can be retrieved reasonably well by fitting only the anisotropy parameter while the spectral index remains (almost) constant for all simulations.
The agreement of the model and the simulation results also supports the claim 
that the waves at small wave numbers dominate the interactions with the 
particles.
Otherwise, the spectral index $s$ would have to be changed according to the 
particle energy.
The low energy particles, e.g.~$100 \, \mathrm{keV}$, resonate with plasma 
waves in the high wave number regime, where the spectrum is steeper.
Thus, according to 
Figs.~\ref{fig:turbulence_production_2d_spectrum_perpendicular} 
and~\ref{fig:turbulence_production_2d_spectrum_parallel}, the 
effective spectral index $s$ should increase for these particles, if the 
resonant interactions with high-$k$ waves were important.
However, this seems not to be the case.
%One could also argue that the model for fast mode turbulence, which is also 
%described by \cite{shalchi_2006}, should be applied in the case of resonant 
%interaction with waves at high perpendicular wave numbers.
But as the results in Figs.~\ref{fig:particles_Dmumu_set10} 
and~\ref{fig:particles_Dmumu_set11} show, a change of the model equations for 
$D_{\mu\mu}$ is not necessary.

The model only fails for the simulations of $10 \, \mathrm{MeV}$-electrons 
(S1e, S1f, S2e, S2f).
This might already be expected from the considerations discussed above:
The high energy electrons are able to resonate with the initially excited 
waves. These waves contain the most energy and thus dominate the interaction 
of the particles with the turbulent spectrum. However, the initial waves 
cannot be considered to be part of the power law 
spectrum itself. As 
Figs.~\ref{fig:turbulence_production_2d_spectrum_perpendicular} 
and~\ref{fig:turbulence_production_2d_spectrum_parallel} show, the energy 
distribution forms a plateau at smallest wave numbers, where the initial 
waves are located. The initial waves are also only few in number, thus not 
forming a continuous spectrum\footnote{Representing a continuous spectrum on 
a discretized grid is always doomed to failure, but at larger wave numbers 
the higher number of individual waves at least creates a rudimentary 
approximation of a continuum.}, but a population of distinct, individual waves.
Thus, the whole model assumption, i.e.~a continuous power law spectrum, is 
invalid. As Fig.~\ref{fig:particles_resonance} shows, the pitch angle diffusion 
coefficient derived from the simulation data can be described reasonably well 
by individual resonances with a number of waves.

Finally, it is worth taking a look at simulations S1f and S2f, which have not 
been discussed so far. These simulations, which employ the same test electron 
energies as S1e and S2e, were carried out to test whether the initial 
particle distribution $f(\mu)$ has an influence on the resulting $D_{\mu\mu}$.
It was already discussed above that the statistical fluctuations tend to 
become more noticeable at those $\mu$, where fewer particles are located.
Thus, reversing the slope of the initial parabola should shift the dominant 
influence of statistical fluctuations from positive $\mu$ to negative.

\begin{figure}[htb]
	\centering
	\includegraphics[width=\linewidth]{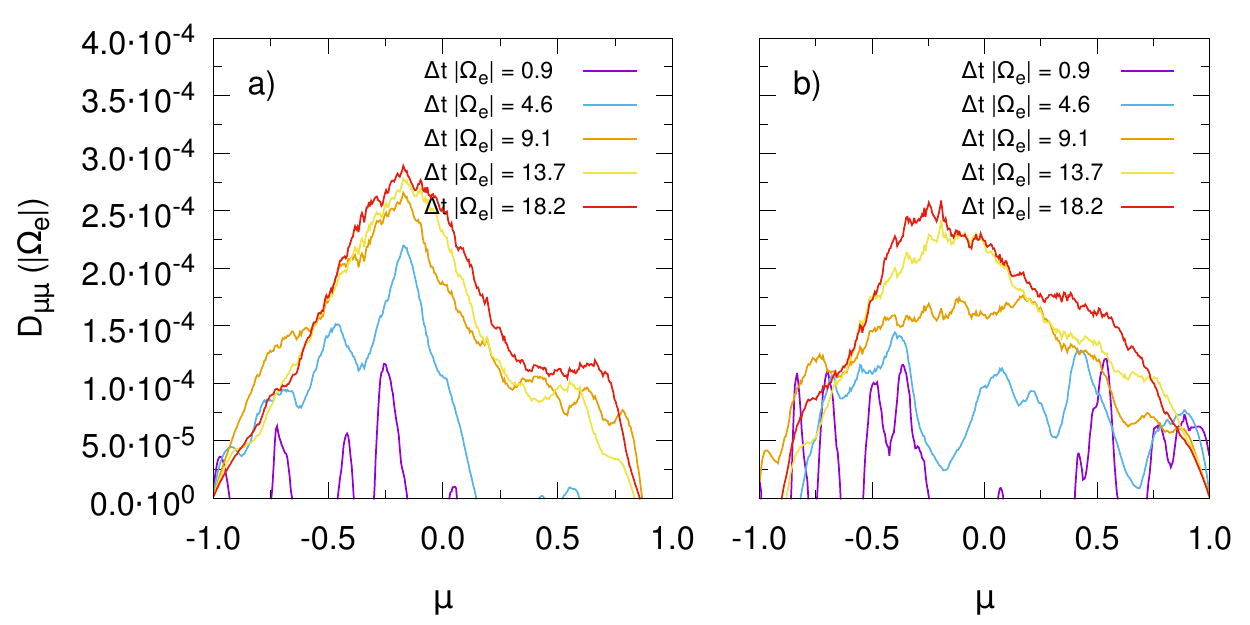}
	\caption{
	Comparison of the pitch angle diffusion coefficients $D_{\mu\mu}$ derived 
from the test electron data of simulations S2e (left) and S2f (right).	The 
two simulations differ by the slope of the parabolic particle distribution $f(
\mu)$ used to initialize the test electrons. The asymmetry of $D_{\mu\mu}$ 
in S2e at early times is not reproduced by S2f, which suggests a numerical or statistical reason for the asymmetry.
	At late times the $D_{\mu\mu}$ become similar, with a single peak near $\mu 
= 0$ in both simulations.
	}
	\label{fig:particles_parabola_11}
\end{figure}

Figure~\ref{fig:particles_parabola_11} depicts the pitch angle diffusion coefficients derived from simulations S2e and S2f in panels a) and b), respectively
%\footnote{A similar figure depicting the comparison of S1e and S1f can be found in Appendix~\ref{app:turbulence_particles}.
%}.
This example is chosen because physical results for $D_{\mu\mu}$ are only 
obtained at negative $\mu$ at early times in S2e.
Should this also be the case in S2f, this would mean that some physical 
process prefers the interaction of waves and energetic particles which 
propagate opposite to the background magnetic field.
However, as Fig.~\ref{fig:particles_parabola_11} shows, this is not the case.
The pitch angle diffusion coefficient derived from S2f appears to be more 
symmetric about $\mu = 0$ at early times.

At late times both simulations produce a single peak in $D_{\mu\mu}$, which 
is located near $\mu = 0$. The peak is slightly shifted to negative $\mu$ in 
both S2e and S2f. This may hint at a physical process leading to the peak not 
being centered exactly around $\mu = 0$. Such an asymmetry is sometimes 
predicted in theoretical models e.g. Schlickeiser \cite{schlickeiser_1989}. 
However, considering the results of simulations of energetic particles and 
their interaction with individual waves, an asymmetry is not expected here.

Thus, the results of simulations S2e and S2f depicted in 
Fig.~\ref{fig:particles_parabola_11} do not entirely agree with the 
expectations. It might be worthwhile to investigate the behavior of the pitch 
angle diffusion coefficient in more detail in a future project.
Changing the initial particle distribution $f(\mu)$ once more (e.g.~by 
altering the parameters $A$, $B$, and $C$ in 
Eq.~(\ref{eq:pitch_angle_parabola})) or reversing the direction of the 
integration over $\mu$ in the method of 
\cite{ivascenko_2016} might help to distinguish between a physically 
motivated asymmetry and numerical artifacts.

\subsection{Conclusion}
\label{sec:results_turbulence_resume}

We have derived a set of pitch-angle diffusion coefficients for dispersive 
whistler waves. Using a Particle-in-Cell code turbulence in the dispersive 
regime has been simulated. Test particle electrons have been injected into 
the simulated turbulence and their transport parameters have been derived.

The conducted turbulence simulations yield power law spectra of the magnetic 
field energy in wave number space. The measured spectral indices are in 
agreement with the findings of \cite{gary_2008,gary_2012}.
Numerical noise limits the energy spectra at high wave numbers, thus 
hindering the production of an energy cascade in the dissipation range.

While the theory is limited to parallel waves, simulations have been performed in two-dimensional wave number space. The theoretical description of oblique, dispersive waves is not practically doable, while one-dimensional turbulence simulations are not producing an energy cascade. The approximation of a parallel spectrum makes this difference between dimensionalities reasonable.

The simulations of energetic particle transport in kinetic turbulence have 
shown that the steep energy spectrum leads to wave-particle interactions 
primarily in the low wave number regime. While low energy particles, in 
principle, resonate with waves in the dispersive or dissipative regime of the 
turbulent cascade, these interactions are subordinate to the interactions 
with non-resonant waves at lower wave numbers.
The reason for this is that the energy content of dispersive waves decreases 
rapidly with increasing wave number, due to the steep power law spectrum.
Thus, the waves at low wave numbers dominate the spectrum as far as particle 
transport is concerned.

This can be seen when comparing simulation data to the theoretical model.
The test electron data from the simulations allows to derive pitch angle 
diffusion coefficients $D_{\mu\mu}$ using the method of \cite{ivascenko_2016}.
Our model for $D_{\mu\mu}$ in plasma turbulence with dispersive waves allows 
for  the prediction of pitch angle diffusion coefficient for Alfv\'en  and 
whistler turbulence.

Simulation data and model match rather well for low-energy electrons. 
Contributions at $\mu=0$ are not modeled correctly as is expected for a 
quasi-linear model. The cross-helicity assumed in model parameters may not 
necessarily represent the cross-helicity of the plasma, but may be to some 
degree a numerical artifact. At higher electron energies particles interact 
with the small number of excited plasma waves which are used as a seed 
population for the generation of kinetic turbulence. The resulting $D_{\mu\mu}$
does not match the prediction for the interaction with the (continuous) 
turbulent spectrum, but can be explained by resonant scattering with several 
waves at discrete wave numbers.

In general simulations of dispersive whistler turbulence and the 
corresponding particle transport are possible, but are also still too 
expensive in terms of computing ressources.

%Thus, while in principle interesting, the effect of dispersion and 
%dissipation on particle transport becomes effectively unimportant in the 
%turbulent medium. The transport of energetic particles therefore depends 
%mostly on the composition of the turbulent spectrum at small wave numbers.

%
%% The MDPI table float is called specialtable
%\begin{specialtable}[H] 
%\small
%\caption{This is a table caption. Tables should be placed in the main text near to the first time they are~cited.\label{tab1}}
%\begin{tabular}{ccc}
%\toprule
%\textbf{Title 1}	& \textbf{Title 2}	& \textbf{Title 3}\\
%\midrule
%Entry 1		& Data			& Data\\
%Entry 2		& Data			& Data\\
%\bottomrule
%\end{tabular}
%\end{specialtable}

%\begin{listing}[H]
%\caption{Title of the listing}
%\rule{\columnwidth}{1pt}
%\raggedright Text of the listing. In font size footnotesize, small, or normalsize. Preferred format: left aligned and single spaced. Preferred border format: top border line and bottom border line.
%\rule{\columnwidth}{1pt}
%\end{listing}

%%%%%%%%%%%%%%%%%%%%%%%%%%%%%%%%%%%%%%%%%%
\authorcontributions{Conceptualization, F.S., C.S. and R.S.; methodology, F.S. and R.S.; software, C.S.; writing---original draft preparation, F.S., C.S. and R.S.; writing---review and editing, F.S., C.S. and R.S.; visualization, F.S. and C.S.;   funding acquisition, F.S. All authors have read and agreed to the published version of the manuscript.}

\funding{This research was funded by Deutsche Forschungsgemeinschaft (DFG) through Grant No. SP 1124/9.}

\acknowledgments{F.S. would like to thank the Deutsche Forschungsgemeinschaft (DFG) for support through Grant No. SP 1124/9. 

The authors gratefully acknowledge the data storage service SDS@hd supported by the 
Ministry of Science, Research and the Arts Baden-Württemberg (MWK) and the
German Research Foundation (DFG) through grant INST 35/1314-1 FUGG and INST 35/1503-1 FUGG.

 The authors gratefully acknowledge the Gauss Centre for Supercomputing e.V. (www.gauss-centre.eu) for funding this project by providing computing time on the GCS Supercomputer SuperMUC at Leibniz Supercomputing Centre (www.lrz.de) through grant pr84ti. 

We acknowledge the use of the \emph{ACRONYM} code and would like to thank the developers (Verein zur F\"orderung kinetischer Plasmasimulationen e.V.) for their support.}

\conflictsofinterest{The authors declare no conflict of interest.} 

\end{paracol}
%%%%%%%%%%%%%%%%%%%%%%%%%%%%%%%%%%%%%%%%%%
% To add notes in main text, please use \endnote{} and un-comment the codes below.
%\begin{adjustwidth}{-5.0cm}{0cm}
%\printendnotes[custom]
%\end{adjustwidth}
%%%%%%%%%%%%%%%%%%%%%%%%%%%%%%%%%%%%%%%%%%
\reftitle{References}

\bibliography{sample}

% If authors have biography, please use the format below
%\section*{Short Biography of Authors}
%\bio
%{\raisebox{-0.35cm}{\includegraphics[width=3.5cm,height=5.3cm,clip,keepaspectratio]{Definitions/author1.pdf}}}
%{\textbf{Firstname Lastname} Biography of first author}
%
%\bio
%{\raisebox{-0.35cm}{\includegraphics[width=3.5cm,height=5.3cm,clip,keepaspectratio]{Definitions/author2.jpg}}}
%{\textbf{Firstname Lastname} Biography of second author}

% The following MDPI journals use author-date citation: Admsci,  Arts, Econometrics, Economies, Genealogy, Humanities, IJFS, Jintelligence, JRFM, Languages, Laws, Literature, Religions, Risks, Social Sciences. For those journals, please follow the formatting guidelines on http://www.mdpi.com/authors/references
% To cite two works by the same author: \citeauthor{ref-journal-1a} (\citeyear{ref-journal-1a}, \citeyear{ref-journal-1b}). This produces: Whittaker (1967, 1975)
% To cite two works by the same author with specific pages: \citeauthor{ref-journal-3a} (\citeyear{ref-journal-3a}, p. 328; \citeyear{ref-journal-3b}, p.475). This produces: Wong (1999, p. 328; 2000, p. 475)

%%%%%%%%%%%%%%%%%%%%%%%%%%%%%%%%%%%%%%%%%%
%% for journal Sci
%\reviewreports{\\
%Reviewer 1 comments and authors’ response\\
%Reviewer 2 comments and authors’ response\\
%Reviewer 3 comments and authors’ response
%}
%%%%%%%%%%%%%%%%%%%%%%%%%%%%%%%%%%%%%%%%%%
\end{document}